\documentclass[%
 reprint,
superscriptaddress,
 amsmath,amssymb,
 aps, prb,
floatfix,
]{revtex4-1}
\usepackage[normalem]{ulem}
\usepackage{graphicx}
\usepackage{dcolumn}
\usepackage{bm}
\usepackage{color}
\begin{document}


\title{Strain and composition dependencies of the near bandgap optical transitions in monoclinic (Al$_x$Ga$_{1-x}$)$_2$O$_3$ alloys with coherent biaxial in-plane strain on (010) Ga$_2$O$_3$}

\author{Rafa\l{} Korlacki}
\email{rkorlacki2@unl.edu}
\affiliation{Department of Electrical and Computer Engineering, University of Nebraska-Lincoln, Lincoln, NE 68588, USA}
\author{Matthew Hilfiker}
\affiliation{Department of Electrical and Computer Engineering, University of Nebraska-Lincoln, Lincoln, NE 68588, USA}
\author{Jenna Knudtson}
\affiliation{Department of Electrical and Computer Engineering, University of Nebraska-Lincoln, Lincoln, NE 68588, USA}
\author{Megan Stokey}
\affiliation{Department of Electrical and Computer Engineering, University of Nebraska-Lincoln, Lincoln, NE 68588, USA}
\author{Ufuk Kilic}
\affiliation{Department of Electrical and Computer Engineering, University of Nebraska-Lincoln, Lincoln, NE 68588, USA}
\author{Akhil Mauze}
\affiliation{Materials Department, University of California Santa Barbara, Santa Barbara, CA 93106, USA}
\author{Yuewei Zhang}
\affiliation{Materials Department, University of California Santa Barbara, Santa Barbara, CA 93106, USA}
\author{James Speck}
\affiliation{Materials Department, University of California Santa Barbara, Santa Barbara, CA 93106, USA}
\author{Vanya Darakchieva}
\affiliation{Terahertz Materials Analysis Center and Center for III-N technology, C3NiT -- Janz\`{e}n, Department of Physics, Chemistry and Biology (IFM), Link\"{o}ping University, 58183 Link\"{o}ping, Sweden}
\affiliation{NanoLund and Solid State Physics, Lund University, 22100 Lund, Sweden}
\author{Mathias Schubert}
\email{schubert@engr.unl.edu}
\homepage{http://ellipsometry.unl.edu}
\affiliation{Department of Electrical and Computer Engineering, University of Nebraska-Lincoln, Lincoln, NE 68588, USA}
\affiliation{Terahertz Materials Analysis Center and Center for III-N technology, C3NiT -- Janz\`{e}n, Department of Physics, Chemistry and Biology (IFM), Link\"{o}ping University, 58183 Link\"{o}ping, Sweden}

\date{\today}

\begin{abstract}
The bowing of the energy of the three lowest band-to-band transitions in $\beta$-(Al$_{x}$Ga$_{1-x}$)$_2$O$_3$ alloys was resolved using a combined density functional theory (DFT) and generalized spectroscopic ellipsometry (GSE) approach. The DFT calculations of the electronic band structure of both, $\beta$-Ga$_2$O$_3$ and $\theta$-Al$_2$O$_3$, allow extracting of the linear portion of the energy shift in the alloys, and provide a method for quantifying the role of coherent strain present in the $\beta$-(Al$_{x}$Ga$_{1-x}$)$_2$O$_3$ thin films on (010) $\beta$-Ga$_2$O$_3$ substrates. The energies of band-to-band transitions were obtained using the spectroscopic ellipsometry eigenpolarization model approach [A. Mock et al., Phys. Rev. B 95, 165202 (2017)]. After subtracting the effects of strain which also induces additional bowing and after subtraction of the linear portion of the energy shift due to alloying, the bowing parameters associated with the three lowest band-to-band transitions in monoclinic $\beta$-(Al$_{x}$Ga$_{1-x}$)$_2$O$_3$ are found. 
\end{abstract}

\maketitle

\section{Introduction}


The ternary alloy (Al$_x$Ga$_{1-x}$)$_2$O$_3$ of gallia ($\beta$-Ga$_2$O$_3$, bandgap E$_g$=5.04~eV)\cite{Mock_2017Ga2O3} and alumina ($\alpha$-Al$_2$O$_3$, E$_g$=9.2~eV)\cite{doi:10.1063/1.357922} is of current interest for applications in high power electronic devices. The critical electric field in semiconductor electronic devices is proportional to the bandgap energy with the power of 1.83.\cite{Slobodyan2022} Hence, alloys of gallium oxide and aluminum oxide permit for device architectures potentially with very large breakdown fields, and offer a tunable ultrawide bandgap reaching far into the ultraviolet-C spectral region.\cite{doi:10.1063/5.0078037}  Accurate characterization and understanding of the bandgap properties are prerequisites for future device designs.

\begin{figure}[!tbp]
  \begin{center}
    \includegraphics[width=.75\linewidth]{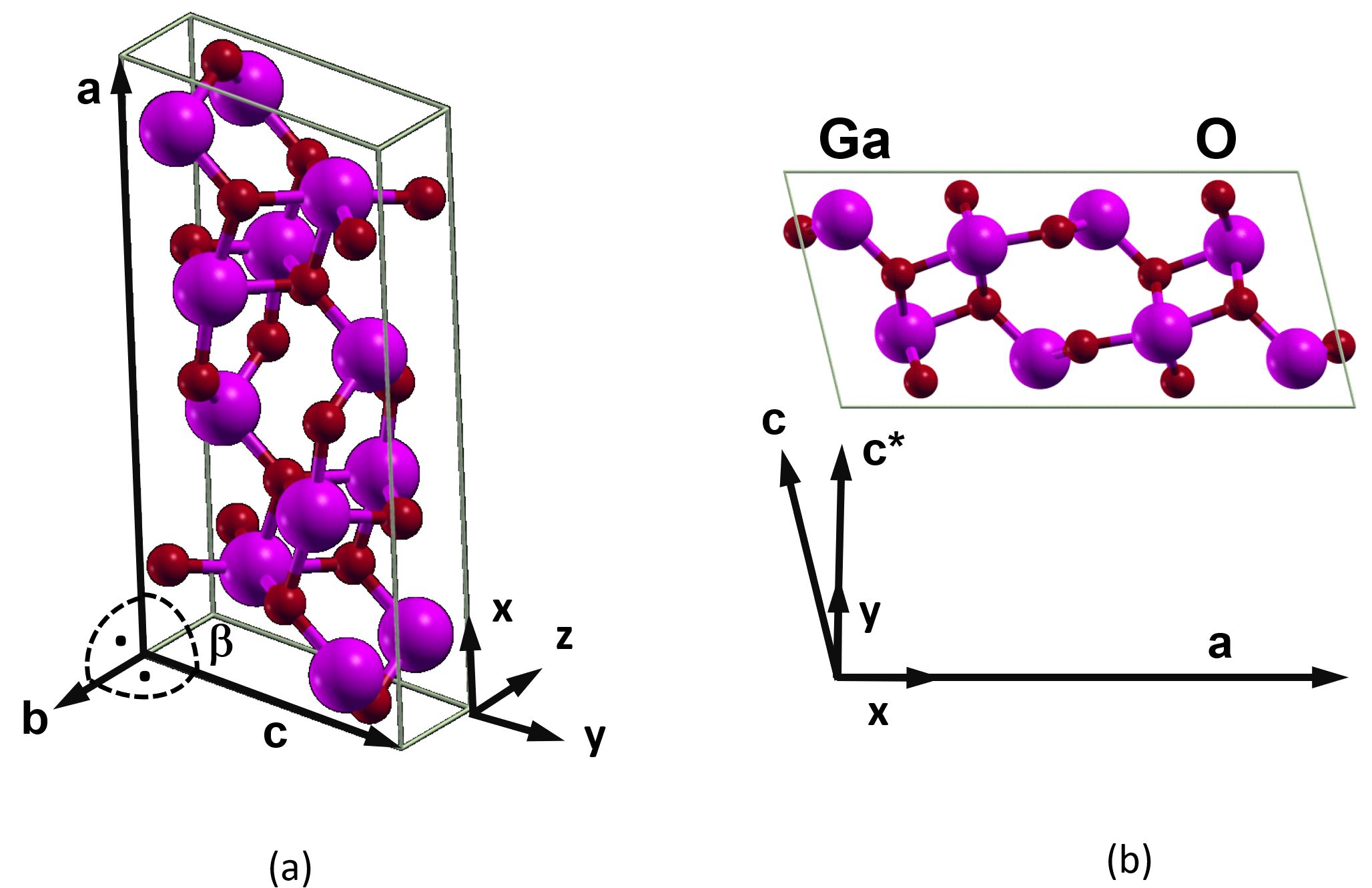}
    \caption{(a) Definition of Cartesian laboratory coordinate system ($x$, $y$, $z$), and as an example the unit cell of $\beta$-Ga$_2$O$_3$ with monoclinic angle $\beta$, and crystal unit axes $\mathbf{a}$, $\mathbf{b}$, $\mathbf{c}$. (b) Monoclinic plane $\mathbf{a}$ - $\mathbf{c}$ viewed along axis $\mathbf{b}$. ($\mathbf{b}$ points into the plane.) Vector $\mathbf{c^{\star}}$ parallel to axis $y$ is used for convenience. Reprinted from Ref.~\onlinecite{SchubertPRB2016} with copyright permission by American Physical Society.}
    \label{fig:Ga2O3unitcell}
  \end{center}
\end{figure}

The oxide system (Al,Ga)$_2$O$_3$ exhibits an interesting phase diagram with a number of polymorphs of different symmetries.\cite{PeelaersAlGO} On the gallium oxide side, the monoclinic $\beta$ phase (space group 12) is stable and relatively easy to grow as a bulk crystal. The unit cell is shown in Fig.~\ref{fig:Ga2O3unitcell}. The rhombohedral $\alpha$ phase (space group 167) is metastable and has been synthesized epitaxially.\cite{Shinohara_2008} On the aluminum oxide side the rhombohedral $\alpha$ phase is stable. The $\alpha$ phase of aluminum oxide occurs naturally as the mineral sapphire. The monoclinic polymorph $\theta$-Al$_2$O$_3$, on the other hand, is metastable and its properties are mostly unknown. Within the ternary alloys the monoclinic structure is expected to be stable and energetically favored up to about 70\% of aluminum concentration.\cite{PeelaersAlGO} The $\alpha$ phase, on the other hand, has been demonstrated to exist  via epitaxial growth throughout the entire composition range with high crystal quality.\cite{doi:10.1126/sciadv.abd5891}


Harman~\textit{et al.}\cite{doi:10.1063/1.357922} provided a value for E$_g$=9.2~eV of $\alpha$-Al$_2$O$_3$ not further discriminating between polarization directions and ignoring excitonic contributions. Accurate analysis of the sapphire bandgap, its anisotropy, and excitonic contributions have not been reported yet. Little information is available for the bandgap properties of the obscure monoclinic phase $\theta$-Al$_2$O$_3$.\cite{Wakabayashi_2021} The bandgap energy and direction dependence for rhombohedral structure $\alpha$-Ga$_2$O$_3$ was recently discussed by Hilfiker~\textit{et al.} from a detailed spectroscopic ellipsometry and density functional theory (DFT) analyses.\cite{doi:10.1063/5.0031424} There, two lowest-energy bandgap transitions were identified with polarization parallel and perpendicular to the lattice $c$ axis in the rhombohedral structure. In $\beta$-Ga$_2$O$_3$ three lowest-energy direct optical transitions at the Brillouin zone center possess different transition energies and are polarized along different directions within the monoclinic unit cell.\cite{Sturm_2016,Mock_2017Ga2O3,Ratnaparkhe_2017} Recently, a generalized spectroscopic ellipsometry analysis revealed the lowest-energy transition between the lowest Brillouin zone center conduction and the highest valence band (transition labeled $\Gamma_{1-1}$) is polarized within the monoclinic plane, and which was determined to occur at a photon energy of 5.04~eV. The onset of absorption caused by exciton formation was found at the lower photon energy of the bandgap reduced by 120~meV. The second-lowest transition ($\Gamma_{1-2}$) also polarized within the monoclinic plane was determined to be 5.40~eV with its excitonic onset of absorption at 5.17~eV. The third-lowest band-to-band transition ($\Gamma_{1-4}$) and excitonic onset of absorption, respectively, were recorded at 5.64~eV and 5.46~eV.\cite{Mock_2017Ga2O3,doi:10.1063/5.0078037} As these transitions involve different pairs of energy bands, and different transition energies, the edge of the direct bandgap appears slightly different in absorption spectra recorded at different polarization and propagation directions relative to the monoclinic crystal axes. Hence, $\beta$-Ga$_2$O$_3$ reveals pleochroism in the ultraviolet-A spectral range. This complex anisotropy and bandgap energy behavior evolves within the alloys of monoclinic structure (Al$_x$Ga$_{1-x}$)$_2$O$_3$ and requires detailed discussions. In this paper, we will focus on the three lowest-energy band-to-band transitions and their evolution with composition $x$ in the monoclinic (Al$_x$Ga$_{1-x}$)$_2$O$_3$. 


Lattice strain affects electronic properties such as eigenvalues and eigenstates. The dependence of the bandgap on strain is often expressed in perturbation theory considerations where small changes to the lattice unit cells cause changes in electronic properties. When such changes are observed to be sufficiently well approximated linear in their causes, i.e., the strain or stress tensor elements, the coefficients rendering analytic expressions between bandgap values and strain or stress parameters are known as strain or stress deformation potential parameters, respectively. Such fist-order deformation potential parameters are very useful in predicting the influence of strain onto bandgap shifts but also for effects on thermal or electrical transport properties, for example. We have recently derived the general strain-stress relationships in linear perturbation theory for monoclinic structure $\beta$-Ga$_2$O$_3$.\cite{Korlacki2020} Using DFT calculations we provided deformation potential parameters for Brillouin zone center phonon modes\cite{Korlacki2020} as well as for the lowest conduction band level and the top valence band energy levels involved in the three lowest-energy band-to-band transitions.\cite{Korlacki_2022} We provide the latter parameters also for monoclinic structure $\theta$-Al$_2$O$_3$ in this work, and use this set or parameters for numerical reduction of the effect of strain from the anisotropic bandgap values observed from the set of strained $\beta$-(Al$_x$Ga$_{1-x}$)$_2$O$_3$ samples investigated in this work by spectroscopic ellipsometry.  


Heteroepitaxial growth of alloys as thin films often results in strained epitaxial material. The surface onto which the epitaxial layer is deposited -- the template -- is characterized by the surface orientation under which a given single crystalline substrate is cut. Under pseudomorphic growth the epitaxial thin film adopts the in-plane (interfacial) lattice parameters of the substrate. The thin film lattice dimension perpendicular to the interface then adjusts according to its elastic properties. In case of pure elastic thin film deformation and in the absence of any defect or partial relaxation, the thin film is then fully strained along the interface (coherent biaxial in-plane strain), and stress free perpendicular to its interface. If the substrate is one of the binary compounds, e.g., $\beta$-Ga$_2$O$_3$, then with increasing composition $x$ the misfit between the template and the epitaxial layer $\beta$-(Al$_x$Ga$_{1-x}$)$_2$O$_3$ enlarges and the magnitude of the coherent biaxial in-plane strain increases. Because Al has a smaller radius than Ga, as a rule of thumb, the in-plane strain will be negative, while the out-of-plane strain will be positive. The strain can be evaluated by measuring x-ray diffraction reciprocal space maps combining analyses of both symmetric and asymmetric reflections and by employing Vegard's rule. This rule assumes that in a ternary system a given lattice property, $P\{x\}$ can be approximated via a linear shift between the properties of the binary endpoints, $P_0=P\{x=0\}$ and $P_1=P\{x=1\}$, $P\{x\}=P_0(1-x)+P_1x$. Combining x-ray diffraction and Vegard's rule can also be used to determine the actual composition of a given film. With the in-plane lattice parameters known, the out of plane lattice parameter can be calculated using the elastic constants. The latter can also be calculated using Vegard's rule from the constants for the binary compounds. It is clear that biaxial in-plane strain will differ for growth on different templates. In this work, we discuss in detail the effects of coherent biaxial in-plane strain with the (010) plane of $\beta$-Ga$_2$O$_3$ as the interfacial plane. Our analysis details can be easily adopted to any other coherent biaxial in-plane strain for as long as the resulting lattice distortion maintains the monoclinic symmetry of the epitaxial layer.     


Typically, in alloys the composition dependence of the bandgap shows a non-linear behavior which is often approximated by a second-order composition dependence also known as bandgap bowing.\cite{PeelaersAlGO} The evolution of the bandgap for a ternary system such as $\beta$-(Al$_x$Ga$_{1-x}$)$_2$O$_3$ also depends on the state of strain which in turn depends on the growth condition, for example, in coherent biaxial in-plane strain growth, the strain depends linearly on the composition $x$. It is important to accurately determine the strain-free composition dependence of the bandgap as well as the influence of strain as a function of composition. We note that coherent biaxial in-plane strain due to composition-induced template mismatch causes additional bandgap bowing as we will discuss further below. In addition to composition and strain, effects such as the degree of atomic ordering, e.g., among equally grouped anionic or cationic sites in polar lattices can influence the composition dependence introducing additional bowing.\cite{PhysRevB.44.7947,PhysRevB.49.14337} Other order-disorder phenomena such as defects may even further affect bowing.\cite{doi:10.1063/1.372014} To determine the ``pure'' composition dependence of the bandgap would only be possible for strain free alloys with either perfect atomic order or perfect atomic disorder and in the absence of any defects and impurities. The latter is unrealistic for experimental conditions, however, one may consider high-quality material and assume perfect atomic disorder in the absence of indications for superstructure ordering in electron microscopy investigations. Then, one may ignore the effects of ordering and defects on the bandgap bowing. If the effects of strain are known, one may then access the ``pure'' bandgap composition dependence by reducing numerically the shifts induced by strain from measured bandgap parameters.     


A complete set of bowing parameters for the two lowest anisotropic bandgap energy parameters was determined by Hilfiker~\textit{et al.} for $\alpha$ phase (Al$_x$Ga$_{1-x}$)$_2$O$_3$ ($0\le x \le 1$) using generalized ellipsometry and DFT analyses.\cite{HilfikeraAGOEg2022} The epitaxial layers were grown on sapphire and free of strain. As an important detail for $\alpha$-(Al$_x$Ga$_{1-x}$)$_2$O$_3$ the character of the valence band structure changes between the two binary compounds. A switch in band order at the top of the valence bands was identified at approximately 40~$\%$ Al content where the lowest band-to-band transition occurs with polarization perpendicular to lattice $c$ direction in $\alpha$-Ga$_2$O$_3$ whereas for $\alpha$-Al$_2$O$_3$ the lowest transition occurs with polarization parallel to lattice $c$ direction. Hence, two bowing parameters are needed when describing the lowest-energy transition throughout the composition range.\cite{HilfikeraAGOEg2022} Additionally, the character of the lowest energy critical point transition for polarization parallel to $\mathbf{c}$ changes from a Van Hove singularity with $M_{1}$ type in $\alpha$-Ga$_2$O$_3$ to $M_{0}$ type van Hove singularity in $\alpha$-Al$_2$O$_3$.\cite{HilfikeraAGOEg2022} Hilfiker~\textit{et al.} also investigated the lowest band-to-band transitions in $\beta$-(Al$_x$Ga$_{1-x}$)$_2$O$_3$ for compositions $x$ up to 21~$\%$.\cite{Hilfiker_2019} The epitaxial layers were grown fully strained on (010) templates. The composition dependencies of the three lowest band-to-band transitions and their excitonic contributions were deduced while any strain influences were ignored. At that time, no information on the effect of strain in $\beta$-(Al$_x$Ga$_{1-x}$)$_2$O$_3$ was available in the literature. No attempt was made to derive bowing parameters because the effect of strain was unknown. Other previous works have reported the lowest-energy onset of absorption in $\beta$-(Al$_x$Ga$_{1-x}$)$_2$O$_3$ as a function of composition $x$, and various reports were given for bowing parameters (see Table~\ref{tab:bowing_lit}). However, except for Hilfiker~\textit{et al.} none of the previous reports considered the peculiar anisotropy and band order in $\beta$-(Al$_x$Ga$_{1-x}$)$_2$O$_3$ and the bowing parameters for all three lowest-energy band to band transitions are unknown.  

In the current paper we perform a combined generalized spectroscopic ellipsometry (GSE) and DFT study of the bandgap bowing of epitaxial $\beta$-(Al$_x$Ga$_{1-x}$)$_2$O$_3$ grown on (010) templates for compositions $x$ up to 21~$\%$. GSE is used to determine properties, including exciton parameters, of the near-bandgap direct optical transitions between the valence and conduction bands for compositions between 4.6~$\%$ and 21~$\%$ of aluminum, while the strain effects and the linear slope across the alloys are obtained from DFT calculations, respectively. Below we describe the coherent biaxial in-plane strain in the specific circumstance of the (010) coherent biaxial in-plane strain, then we detail our first principles calculations followed by the description of ellipsometry measurements, and finally we present and discuss our results.

\begin{table}
\centering
\caption{Previously published values of the bowing parameter in monoclinic $\beta$-(Al$_{x}$Ga$_{1-x}$)$_2$O$_3$.}
\label{tab:bowing_lit}
\begin{ruledtabular}
\begin{tabular}{{l}{c}{c}}
Source & Method & Value (eV)\\
\hline
Kim~\textit{et al.}\cite{Kim_2021} & Calc. & -0.32$^{a}$\\
Bhattacharjee~\textit{et al.}\cite{Bhattacharjee_2021} & Exp. & 0.4\\
Wang~\textit{et al.}\cite{WangAlGODFT} & Calc. & 0.32 ($x<0.5$)$^{a}$\\
& & 0.54 ($x>0.5$)$^{a}$\\
& & 1.0$^{a}$\\
Ratnaparkhe and Lambrecht\cite{Ratnaparkhe_2020} & Calc. & 0.8$^{a}$\\
Peelaers~\textit{et al.}\cite{PeelaersAlGO} & Calc. & 0.93$^{a}$\\
& & 1.37$^{b}$\\
Bhuiyan~\textit{et al.}\cite{Bhuiyan_2020} & Exp. & 1.25$^{c}$\\
Li~\textit{et al.}\cite{Li_2018} & Exp. & 1.3$^{a}$\\
Ota~\textit{et al.}\cite{Ota_2020} & Calc. & 1.4$^{a}$\\
Jesenovec~\textit{et al.}\cite{Jesenovec_2022} & Exp. & 1.69$^{a}$\\
Wakabayashi~\textit{et al.}\cite{Wakabayashi_2021} & Exp. & 2.19 ($x<0.5$)$^{a}$\\
& & 2.27 ($x>0.5$)$^{a}$\\
\end{tabular}
\end{ruledtabular}
$^{a}$ Determined for the indirect bandgap.\\
$^{b}$ Determined for the direct bandgap.\\
$^{c}$ Determined for the conduction band minimum.
\end{table}

\section{Theory}\label{sec:theory}

\subsection{In-plane Strain Parameters}
\label{sec:strainparameters}

The volume of the unit cell of Al$_2$O$_3$ is smaller than that of Ga$_2$O$_3$. As a result, the incorporation of aluminum into Ga$_2$O$_3$ leads to a dimensional lattice change. In the case of pseudomorphic heteroepitaxial growth when the epitaxial layer adopts the interfacial lattice spacing of the template the epitaxial layer is under strain. As a result, the out-of-plane lattice spacing undergoes elastic changes until it is stress free. X-ray diffraction measurements can be used to verify the thin film lattice parameters. Such measurements are needed to verify that no partial lattice relaxation has occurred, e.g., via defect creation. To calculate the strain tensor elements for the epitaxial layer, we must examine the difference between the relaxed and coherently strained lattice parameters. Kranert~\textit{et al.} determined linear relationships (Vegard's rule) as a function of composition for all lattice parameters in the case of relaxed $\beta$-(Al$_x$Ga$_{1-x}$)$_2$O$_3$.\cite{Kranert_2015} For coherent biaxial in-plane strain, i.e., without any partial lattice relaxation, the state of strain will depend on the compositon as well as the surface orientation of the substrate. Oshima~\textit{et al.} obtained the relationship between the length of $\mathbf{b}$-vector and aluminum concentration for $\beta$-(Al$_{x}$Ga$_{1-x}$)$_2$O$_3$ grown on (010) $\beta$-Ga$_2$O$_3$ substrates.\cite{OshimaAPE2016AGO} In the further analysis we assume the following to be true regarding the lattice parameters of the investigated $\beta$-(Al$_x$Ga$_{1-x}$)$_2$O$_3$ films: (i) due to the coherent growth, the lattice parameters of the films parallel to (010) (i.e., within the monoclinic plane) maintain the values of the substrate; (ii) as a result, perpendicular to (010) the layer contracts as demonstrated by Oshima~\textit{et al.}\cite{OshimaAPE2016AGO} For every composition $x$, we calculate the strain tensor elements from the differences of the lattice parameters between the coherent values described above and the relaxed values determined by Kranert~\textit{et al.}\cite{Kranert_2015} 

A monoclinic lattice may be characterized by lattice vectors $\{ \mathbf{a}, \mathbf{b}, \mathbf{c}\}$, as shown, for example, for $\beta$-Ga$_2$O$_3$ in Fig.~\ref{fig:Ga2O3unitcell}, reproduced from Schubert~\textit{et al.}\cite{SchubertPRB2016} The monoclinic angle is defined here as the obtuse between axis $\mathbf{a}$ and $\mathbf{c}$. Hence,

\begin{equation}
\mathbf{a}=\left(a,0,0\right),\mbox{}\mathbf{b}=\left(0,0,-b\right),\mbox{} \mathbf{c}=\left(c_x,c_y,0\right), \end{equation}

\noindent with coordinates and lengths $a$, $b$, $c=\sqrt{c_x^2+c_y^2}$ accordingly. We aim at making use of results in first order perturbation theory, and we require therefore that any strain or stress leaves the symmetry of the unit cell unchanged, hence, we assume that stress ($\sigma$) and strain ($\epsilon$) tensors have monoclinic symmetry as well

\begin{equation}
\epsilon=
\begin{bmatrix}
 \epsilon_{\mathrm{xx}} & \epsilon_{\mathrm{xy}} & 0\\
 \epsilon_{\mathrm{xy}} & \epsilon_{\mathrm{yy}} & 0\\
 0 & 0 & \epsilon_{\mathrm{zz}}\\

\end{bmatrix},
\end{equation}
\begin{equation}
\sigma=
\begin{bmatrix}
 \sigma_{\mathrm{xx}} & \sigma_{\mathrm{xy}} & 0\\
 \sigma_{\mathrm{xy}} & \sigma_{\mathrm{yy}} & 0\\
 0 & 0 & \sigma_{\mathrm{zz}}\\
\end{bmatrix}.
\end{equation}

\noindent In general, a lattice vector $\mathbf{v}$ under strain changes such as

\begin{equation}
\Delta \mathbf{v}=\mathbf{v'}-\mathbf{v}=\epsilon\mathbf{v}.
\end{equation}

\noindent A state of strain (or stress) maybe expressed by lengths $a'$, $b'$, $c'$, and monoclinic angle $\beta'$. It is further assumed that the monoclinic plane of the strained lattice remains within the $(x,y)$ plane. Hence,

\begin{equation}
\mathbf{a'}=a\left(1+\epsilon_{\mathrm{xx}},\epsilon_{\mathrm{xy}},0\right), 
\end{equation}
\noindent and
\begin{equation}
\mathbf{c'}= \left(c_x[1+\epsilon_{\mathrm{xx}}]+c_y\epsilon_{\mathrm{xy}},c_y[1+\epsilon_{\mathrm{yy}}]+c_x\epsilon_{\mathrm{xy}},0\right).
\end{equation}

\noindent We define $\alpha = 180^{\circ}-\beta$ and $\alpha' = 180^{\circ}-\beta'$. Then

\begin{equation}
\frac{\Delta a}{a}=\frac{a'-a}{a},
\end{equation}
\noindent and
\begin{equation}
\frac{\Delta c_x}{c_x}=\frac{c'_x-c_x}{c_x}=\frac{c'\cos{\alpha'}-c\cos{\alpha}}{c\cos{\alpha}},
\end{equation}
\begin{equation}
\frac{\Delta c_y}{c_y}=\frac{c'_y-c_y}{c_y}=\frac{c'\sin{\alpha'}-c\sin{\alpha}}{c\sin{\alpha}}.
\end{equation}

\noindent A quadratic equation emerges for $\epsilon_{\mathrm{xx}}$ with standard normalized coefficients

\begin{equation}
p=2\left(\frac{1-\frac{\Delta c_x}{c_x}\cot^2{\alpha}}{1+\cot^2{\alpha}}\right),
\end{equation}
\begin{equation}
q=\frac{1+(\frac{\Delta c_x}{c_x})^2\cot^2{\alpha}-(\frac{\Delta a}{a}+1)^2}{1+\cot^2{\alpha}},
\end{equation}
\begin{equation}
\label{eq:epsxx}
\epsilon_{\mathrm{xx}}=-\frac{p}{2} \pm \sqrt{\left(\frac{p}{2}\right)^2-q}.
\end{equation}

\noindent  In the latter equation, the negative branch is discarded as unphysical solution. $\epsilon_{\mathrm{xy}}$, $\epsilon_{\mathrm{yy}}$, $\epsilon_{\mathrm{zz}}$ then follow from simple algebra:

\begin{equation}
\label{eq:epsxy}
\epsilon_{\mathrm{xy}}=\left(\epsilon_{\mathrm{xx}}-\frac{\Delta c_x}{c_x}\right)\cot{\alpha},
\end{equation}
\begin{equation}
\label{eq:epsyy}
\epsilon_{\mathrm{yy}}=\frac{\Delta c_y}{c_y}+\epsilon_{\mathrm{xy}}\cot{\alpha}.
\end{equation}

For the strain parameter along axis $\mathbf{b'}$
\begin{equation}
\mathbf{b'}=b\left(0,0,-\left[1-\epsilon_{\mathrm{zz}}\right]\right), 
\end{equation}
\begin{equation}
\frac{\Delta b}{b}=\frac{b'-b}{b},
\end{equation}
\noindent and
\begin{equation}
\epsilon_{\mathrm{zz}}=\frac{\Delta b}{b}.
\end{equation}

\subsection{First Principles Calculations}

DFT calculations are employed here to quantitatively estimate two independent effects: the linear slope of the composition dependence, i.e., Vegard's rule for the energy of selected band-to-band transitions, and the composition dependence of the effect of strain onto these transitions, respectively. To begin with, we have to verify that the band structures of the monoclinic phases of these two materials, $\beta$-Ga$_2$O$_3$ and $\theta$-Al$_2$O$_3$, are similar such that a one-to-one correspondence between the $\Gamma$-point band-to-band transitions in these two materials is present. Only if such similarity exists between the two binary endpoint one may anticipate the validity of Vegard's rule, and a linear composition dependence may be valid for the strain free alloys. To maintain maximum consistency between calculations for the two different materials, all parameters and procedures are kept the same where possible, except for lattice dimensions and atomic potentials.

The effects of strain onto the near-bandgap optical transitions in $\beta$-Ga$_2$O$_3$ were reported by us recently.\cite{Korlacki_2022} The previous results serve here as our starting point. DFT calculations were performed with a fine Monkhorst-Pack\cite{Monkhorst1976} grid for Brillouin zone sampling ($8 \times 8 \times 8$) and a very large plane-wave basis (400 Ry cutoff). A convergence threshold of $1 \times 10^{-12}$ Ry was used to reach self-consistency. These parameters were selected specifically for studying the effects of strain,\cite{Korlacki2020} and we maintain these parameters in the calculations included here. A convergence threshold of $1 \times 10^{-12}$ Ry was used to reach self-consistency. All DFT calculations were performed using a plane-wave code Quantum ESPRESSO.\cite{[{Quantum ESPRESSO is available from http://www.qu\-an\-tum-es\-pres\-so.org. See also: }]GiannozziJPCM2009QE} The calculations for $\beta$-Ga$_2$O$_3$ in Refs.~\onlinecite{Korlacki2020,Korlacki_2022} used  generalized-gradient-approximation (GGA) density functional of Perdew, Burke, and Ernzerhof\cite{PBE1996} in combination with norm-conserving Troullier-Martins pseudopotentials generated using FHI98PP\cite{FuchsCPC1999,TroullierPRB1991} code and available in the Quantum ESPRESSO pseudopotentials library. The pseudopotential for gallium did not include the semicore $3d$ states in the valence configuration. The exact same calculations were repeated for $\theta$-Al$_2$O$_3$ using a pseudopotential for aluminum from the same family of Fritz-Haber Institute Troullier-Martins pseudopotentials. The equilibrium cell was obtained with the same criteria as used in Ref.~\onlinecite{Korlacki2020} for $\beta$-Ga$_2$O$_3$, i.e., $1 \times 10^{-6}$ Ry for energy and $1 \times 10^{-5}$ Ry/bohr for forces.

\subsubsection{Strain Free Band Structure Comparison}

\begin{figure}[!tbp]
  \begin{center}
    \includegraphics[width=.95\linewidth]{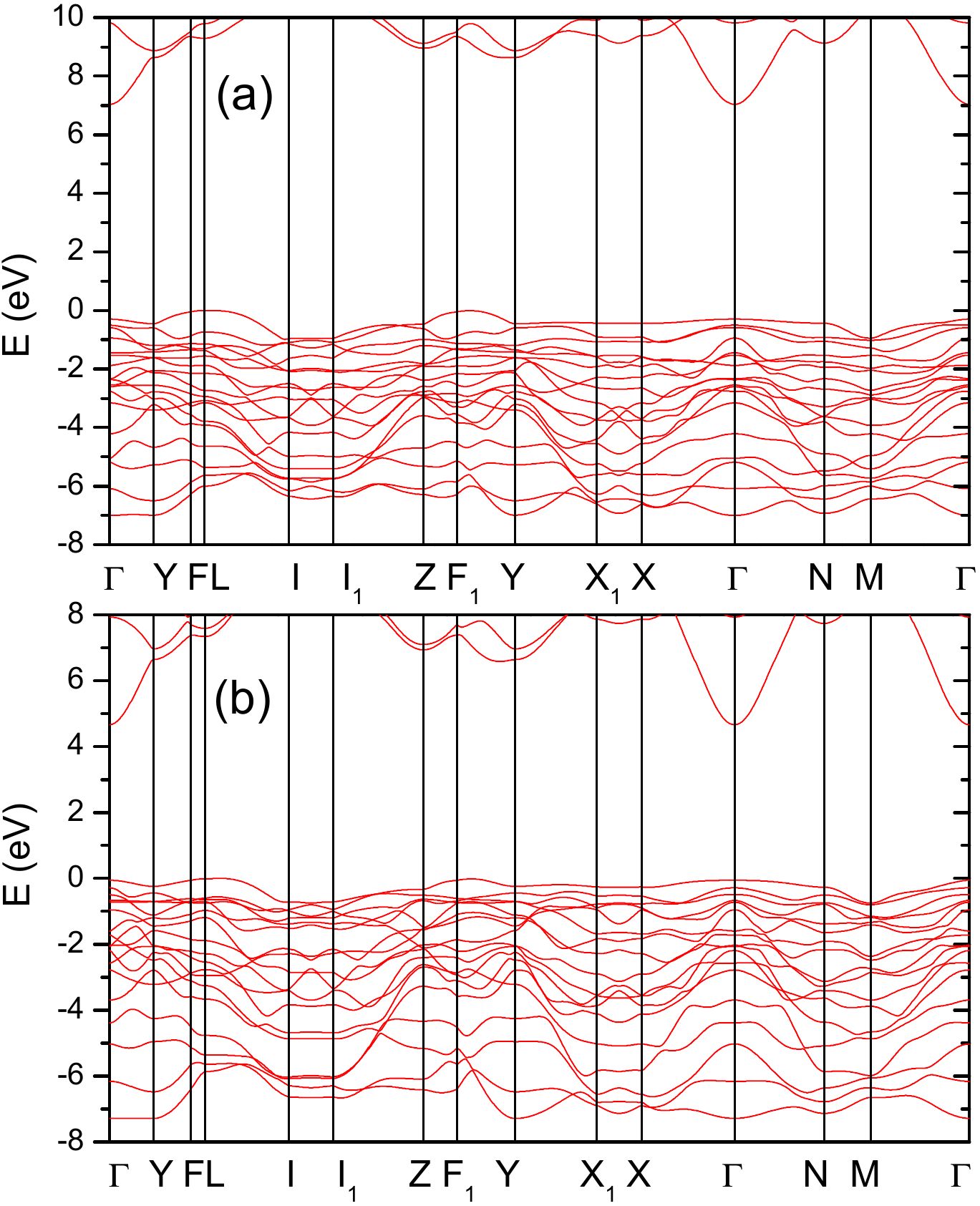}
    \caption{Comparison of the band structure of $\theta$-Al$_2$O$_3$ (a) and the band structure of $\beta$-Ga$_2$O$_3$ (b). Labeling of high symmetry points as in Ref.~\onlinecite{setyawan2010}.}
    \label{fig:band_structure}
  \end{center}
\end{figure}

The analysis of optical matrix elements between the valence and the conduction band at the $\Gamma$-point indicate that $\theta$-Al$_2$O$_3$ indeed exhibits a similar structure of lowest band-to-band transitions than $\beta$-Ga$_2$O$_3$. We consistently use transition labels introduced by Mock~\textit{et al.},\cite{Mock_2017Ga2O3} $\Gamma_{c-v}$, where $c$ and $v$ are indices of the conduction and valence bands, respectively. Numerals begin at the bottom and the top of the conduction and valence bands, respectively, and indicate those participating in a given optical transition. Transitions $\Gamma_{1-1}$ and $\Gamma_{1-2}$ are polarized within the monoclinic plane as valence bands $v=1$ and $v=2$ belong to the irreducible representation $B_{\mathrm{u}}$. The transitions $\Gamma_{1-1}$ has its transition dipole oriented close to the crystallographic vector $\mathbf{c}$, while the transition $\Gamma_{1-2}$ has its transition dipole oriented close to the crystallographic vector $\mathbf{a}$. The transition $\Gamma_{1-4}$ is polarized along the symmetry axis (parallel to the crystallographic vector $\mathbf{b}$).

In order to compare the band structures of $\beta$-Ga$_2$O$_3$ and $\theta$-Al$_2$O$_3$ and obtain reasonably accurate Vegard parameters (slopes) for transition energy parameters, we performed additional band structure calculations using a hybrid Gau-PBE\cite{song2011,song2013} density functional. The hybrid-DFT calculations were performed at the PBE equilibrium geometries and a regular non-shifted $8 \times 8 \times 8$ Monkhorst-Pack grid for the Brillouin zone sampling and $4 \times 4 \times 4$ grid for sampling of the Fock operator. The cutoff for the Fock operator was also reduced to 400 Ry. The convergence threshold for self-consistency in hybrid functional calculations was set to $1 \times 10^{-10}$ Ry. The results of the current calculations for $\beta$-Ga$_2$O$_3$ agree very well with those previously published by Mock~\textit{et al.};\cite{Mock_2017Ga2O3} the previous calculations performed on a coarser $6 \times 6 \times 6$ grid and at slightly different geometry due to a smaller basis (100 Ry). The energies of optical transitions between these two calculations are the same to within 0.02 eV (i.e. $<0.35$\%). The high-resolution interpolated plots of the band structures, shown in Figure~\ref{fig:band_structure}, were obtained with the help of the band interpolation method based on the maximally localized Wannier functions\cite{PhysRevB.56.12847,PhysRevB.65.035109} as implemented in the software package WANNIER90.\cite{mostofi2008} 

\begin{table}
\centering
\caption{Lowest-energy direct bandgap energies and Vegard parameters (slopes) as reported previously and obtained in the present work, for $\beta$-Ga$_2$O$_3$ and  $\theta$-Al$_2$O$_3$.}
\label{tab:slopes}
\begin{ruledtabular}
\begin{tabular}{{l}{c}{c}{c}}
 & $\beta$-Ga$_2$O$_3$ & $\theta$-Al$_2$O$_3$ & Slope \\
 Transition & x=0 (eV) & x=1 (eV) & $E_g$(1) - $E_g$(0) (eV) \\
\hline
$E_g^{a}$ & 4.87 & 7.51 & 2.64\\
$E_g^{b}$ & 4.87 & 7.74 & 2.87\\
$E_g^{c}$ & 4.51 & 6.91 & 2.40 \\
\hline
$\Gamma_{1-1}^{d}$ & 4.725 & 7.337 & 2.613\\
$\Gamma_{1-2}^{d}$ & 4.952 & 7.546 & 2.594\\
$\Gamma_{1-4}^{d}$ & 5.333 & 7.980 & 2.648\\
\end{tabular}
\end{ruledtabular}
$^{a}$Theory (Peelaers~\textit{et al.}~Ref.~\onlinecite{PeelaersAlGO}).\\
$^{b}$Theory (Mu~\textit{et al.}~Ref.~\onlinecite{Mu_2022}).\\ 
$^{c}$Exp., Reflection electron energy loss spectroscopy (Wakabayashi~\textit{et al.}~\onlinecite{Wakabayashi_2021}).\\
$^{d}$Theory (This work).
\end{table}

\subsubsection{Strain and Composition Effects}

The strain effect on the Brillouin zone center energy levels $c=1$ and $\nu = 1,2,4$, hence, on the energies of the band-to-band transitions for $\beta$-Ga$_2$O$_3$ were reported previously.\cite{Korlacki_2022} They were shown to involve four deformation potentials, one associated with each independent component of the monoclinic strain tensor.\cite{Korlacki_2022} Here, we propose a model to account for strain effects in alloys across the entire range of compositions between $\beta$-Ga$_2$O$_3$ and $\theta$-Al$_2$O$_3$. First, we apply the methodology from Refs.~\onlinecite{Korlacki2020,Korlacki_2022} to $\theta$-Al$_2$O$_3$. In these earlier publications, the strain deformation potentials were obtained by testing various deformation patterns (such as hydrostatic pressure, uniaxial stress, and uniaxial strain) and computing energy eigenvalues, i.e., using DFT, for a total of 64 structures representing different deformation scenarios and varying amplitudes of strain. For the determination of the deformation potentials the allowed strain was limited to $\pm 0.0035$, corresponding to 35 out of the 64 strain scenarios. In the current study, we performed analogous DFT calculations for $\theta$-Al$_2$O$_3$. We only performed the calculations for strain patterns within the above $\pm 0.0035$ limit. However, as the cell volume of $\theta$-Al$_2$O$_3$ is smaller than $\beta$-Ga$_2$O$_3$, for the determination of the deformation potentials, we further lowered the strain limit to $\pm 0.0015$. As a result, the deformation potentials were obtained from 21 structures and are listed in Table~\ref{tab:theta_AO}. The obtained deformation potentials accurately reproduce band energies under the 21 deformations included here (Figure~\ref{fig:thAO_strain_fits}). The explicit strain scenarios are given in the supplement of Ref.~\onlinecite{Korlacki2020}.

\begin{table}
\centering
\caption{Linear strain deformation potential parameters in units of eV per unit strain for $\Gamma$-point conduction and valence bands associated with the three lowest band-to-band transitions in $\theta$-Al$_2$O$_3$ calculated using DFT. Note that these strain potentials, and those for $\beta$-Ga$_2$O$_3$ in Ref.~\onlinecite{Korlacki_2022}, describe the strain effects on band energies relative to the energy levels of the strain free equilibrium cell, without considering effects of band offsets. In order to calculate the latter, one needs to augment considerations of bulk (equilibrium) Fermi level shifts as well.}
\label{tab:theta_AO}
\begin{ruledtabular}
\begin{tabular}{{l}{c}{c}{c}{c}}
Band & P$_{\eta, xx}$ & P$_{\eta, xy}$ & P$_{\eta, yy}$ & P$_{\eta, zz}$ \\
\hline  
$c = 1$ & -20.02 & 5.40 & -21.86 & -19.59 \\
$v = 1$ & -10.73 & 4.83 & -14.70 & -8.31 \\
$v = 2$ & -11.63 & 1.57 & -11.91 & -10.84 \\
$v = 4$ & -12.34 & 2.58 & -6.58 & -12.28 \\
\end{tabular}
\end{ruledtabular}
\end{table}

\begin{figure*}[!tbp]
  \begin{center}
    \includegraphics[width=.95\linewidth]{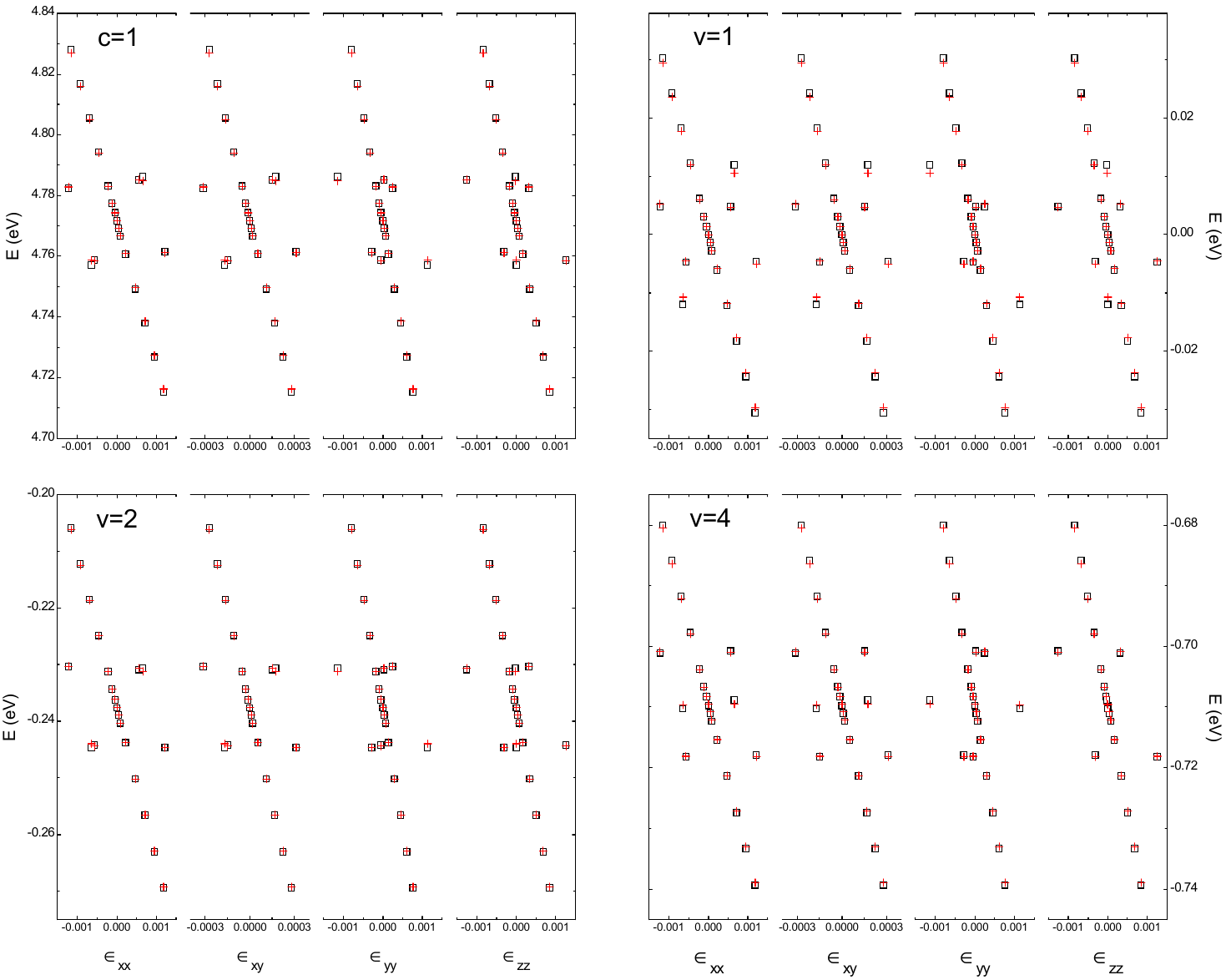}
    \caption{Energies of selected energy bands vs. strain tensor elements $\epsilon_{xx}$; $\epsilon_{xy}$; $\epsilon_{yy}$, and $\epsilon_{zz}$, under various deformations as explained in the text. Squares - energy eigenvalues obtained directly from DFT calculations; crosses - values calculated using deformation potentials from Tab.~\ref{tab:theta_AO}.}
    \label{fig:thAO_strain_fits}
  \end{center}
\end{figure*}

To calculate the effect of strain onto the band energy levels as a function of composition we propose to employ Vegard's rule. Hence, for the deformation potentials for $\beta$-(Al$_x$Ga$_{1-x}$)$_2$O$_3$ we use  values for $\theta$-Al$_2$O$_3$ in Table~\ref{tab:theta_AO} and for $\beta$-Ga$_2$O$_3$ in Table II of Ref.~\onlinecite{Korlacki_2022}. Thus, we propose a set of equations for the combined strain and composition dependencies of the lowest band-to-band transitions ($E_{\Gamma_{1-1}}$, $E_{\Gamma_{1-2}}$, and $E_{\Gamma_{1-4}}$) in $\beta$-(Al$_{x}$Ga$_{1-x}$)$_2$O$_3$:

\begin{equation}
\label{eq:eg11}
\begin{split}
    E_{\Gamma_{1-1}} (\mathrm{eV}) = 5.04+2.613x\\ + (-8.38 - 0.90x) \epsilon_{\mathrm{xx}} + (0.346 + 0.22x)\epsilon_{\mathrm{xy}}\\
    +(-5.93 - 1.23x)\epsilon_{\mathrm{yy}} +(-9.56 -1.71x)\epsilon_{\mathrm{zz}}\\
    +b_{\Gamma_{1-1}}x(x-1),
\end{split}    
\end{equation}

\begin{equation}
\label{eq:eg12}
\begin{split}
    E_{\Gamma_{1-2}} (\mathrm{eV}) = 5.40+2.594x\\ + (-7.75 - 0.63x) \epsilon_{\mathrm{xx}} + (2.29 + 1.54x)\epsilon_{\mathrm{xy}}\\
    +(-8.49 -1.46x) \epsilon_{\mathrm{yy}} +(-7.74 - 1.01x)\epsilon_{\mathrm{zz}}\\
    +b_{\Gamma_{1-2}}x(x-1),
\end{split}
\end{equation}

\begin{equation}
\label{eq:eg14}
\begin{split}
    E_{\Gamma_{1-4}} (\mathrm{eV}) = 5.64+2.648x\\ + (-8.43 + 0.76x) \epsilon_{\mathrm{xx}} + (3.34 - 0.52x) \epsilon_{\mathrm{xy}}\\
    +(-11.9 - 3.41x) \epsilon_{\mathrm{yy}} + (-6.33 - 0.98x)\epsilon_{\mathrm{zz}}\\
    +b_{\Gamma_{1-4}}x(x-1),
\end{split}
\end{equation}

\noindent where $x$ is the aluminum concentration and $\epsilon_{\mathrm{xx}}$, $\epsilon_{\mathrm{xy}}$, $\epsilon_{\mathrm{yy}}$, and $\epsilon_{\mathrm{zz}}$ are the strain tensor elements. The bowing parameters, $b_{\Gamma_{1-1}}$, $b_{\Gamma_{1-2}}$, and $b_{\Gamma_{1-4}}$, are to be determined further below. The strain free transition energies for $x=0$ are experimental values from Ref.~\onlinecite{Mock_2017Ga2O3}, the slope parameters are from Table~\ref{tab:slopes}. We note that, although not instantly obvious, the strain parameters depend linearly on composition $x$ for coherent biaxial in-plane (010) strain. This can be simply verified by numerical calculations using equations derived above in Sect.~\ref{sec:strainparameters}. As a consequence, and because the deformation potentials depend on composition, it is obvious that strain introduces a parabolic dependence on composition and can be mistaken as bowing if not considered appropriately.  

\subsubsection{Elastic tensor}

We augment the DFT calculations of $\theta$-Al$_2$O$_3$ with the calculation of the full tensor of elastic constants with the help of thermo\_pw code.\cite{Footnote2} This calculation was performed using the same parameters as described above. The tensor of elastic constants (in kbar), in the Voigt notation and standard ordering is given below. Note that the standard ordering is not consistent with our choice of the Cartesian reference system, but rather with $y || \textbf{b}$ and $z || \mathbf{c^{\star}}$.

\begin{equation}
C_{Al_2O_3} = 
\begin{bmatrix}
 2648 & 1113 & 1290 & 0 & -273 & 0 \\
 1113 & 3846 & 589 & 0 & 143 & 0 \\
 1290 & 589 & 3996 & 0 & 146 & 0 \\
 0 & 0 & 0 & 697 & 0 & 191 \\
 -273 & 143 & 146 & 0 & 1018 & 0 \\
 0 & 0 & 0 & 191 & 0 & 1220
\end{bmatrix}.
\end{equation}

The values of the bulk modulus, calculated from the relevant components of the elastic (stiffness) and compliance tensors, are: 1831 kbar (Voigt approximation), 1808 kbar (Reuss approximation), and 1819 kbar (Voigt-Reuss-Hill average).

Together with the elastic tensor for $\beta$-Ga$_2$O$_3$, provided in the supplementary material of Ref.~\onlinecite{Korlacki2020}, allows us to once again apply a linear interpolation scheme to obtain a model of the elastic tensor for the $\beta$-(Al$_{x}$Ga$_{1-x}$)$_2$O$_3$ alloys:

\begin{widetext}
\begin{equation}
\label{eq:C_AlGO}
C_{(Al_{x}Ga_{1-x})_2O_3} = 
\begin{bmatrix}
 2143+505x & 1103+10x & 1200+90x & 0 & -197-76x & 0 \\
 1103+10x & 3300+546x & 669-80x & 0 & 119+24x & 0 \\
 1200+90x & 669-80x & 3248+748x & 0 & 75+71x & 0 \\
 0 & 0 & 0 & 500+197x & 0 & 182+9x \\
 -197-76x & 119+24x & 75+71x & 0 & 689+329x & 0 \\
 0 & 0 & 0 & 182+9x & 0 & 949+271x
\end{bmatrix}.
\end{equation}
\end{widetext}

\subsubsection{Coherent biaxial in-plane strain for (010) surface}

Strain elements within the monoclinic plane, $\epsilon_{xx}$, $\epsilon_{xy}$, and $\epsilon_{yy}$, can be calculated using equations~\ref{eq:epsxx}-\ref{eq:epsyy}, from the differences of the relaxed lattice parameters of alloys versus those of the substrate. Two options arrive for determining $\epsilon_{zz}$, either one determines $b'$ from XRD measurements, or one may use the  tensor of elastic constants, $C$, to calculate $\epsilon_{zz}$. The tensors $C$ are discussed in the previous section. The generalized Hooke's law is then  
\begin{equation}
\tilde{\sigma}=C\tilde{\epsilon},   
\end{equation}
\noindent with
\begin{equation}
    \tilde{\sigma}=\left(\sigma_{xx}, \sigma_{zz}, \sigma_{yy},0, \sigma_{xy}, 0\right)^{T},
\end{equation}
\begin{equation}
    \tilde{\epsilon}=\left(\epsilon_{xx}, \epsilon_{zz}, \epsilon_{yy},0, 2\epsilon_{xy}, 0\right)^{T},
\end{equation}
\noindent where $T$ is the transpose operator. Because for (010) coherent biaxial in-plane strain the stress perpendicular to the monoclinic plane is zero ($\sigma_{zz}=0$), one can calculate $\epsilon_{zz}$ after determining $\epsilon_{xx}$, $\epsilon_{xy}$, and $\epsilon_{yy}$
\begin{widetext}
\begin{equation}
    \epsilon_{zz}=-\frac{[(1103+10x)\epsilon_{xx}+2(119+24x)\epsilon_{xy}+(669-80x)\epsilon_{yy}]}{3300+546x}.
\end{equation}
\end{widetext}
\noindent In the above equation we used the elastic tensor from Eq.~\ref{eq:C_AlGO}. 

\section{Experiment}\label{sec:experiment}

\subsection{Samples}\label{sec:samples}

$\beta$-(Al$_x$Ga$_{1-x}$)$_2$O$_3$ epitaxial thin films were deposited on (010) $\beta$-Ga$_2$O$_3$ substrates using plasma-assisted molecular beam epitaxy as described in Ref.~\onlinecite{OshimaAPE2016AGO}. Aluminum  compositions $x$ of 4.6$\%$, 9.7$\%$, 12$\%$, 15$\%$, 16.3$\%$, and 21$\%$ were determined from X-ray diffraction (XRD) measurements as described by Oshima~\textit{et al.}\cite{OshimaAPE2016AGO} All thin film lattice parameters were obtained from x-ray investigations using symmetric and asymmetric reflections as described by Oshima~\textit{et al.} (Ref.~\onlinecite{OshimaAPE2016AGO}). 
Reciprocal space mapping further indicates a fully coherent growth of the alloy thin films on the (010) substrates without any lattice relaxation. Hence, the further modeling of strain effects in the alloy films will be based on the (010) coherent biaxial in-plane strain conditions. 

\subsection{Generalized Spectroscopic Ellipsometry}

\begin{figure}[!tbp]
  \begin{center}
    \includegraphics[width=.95\linewidth]{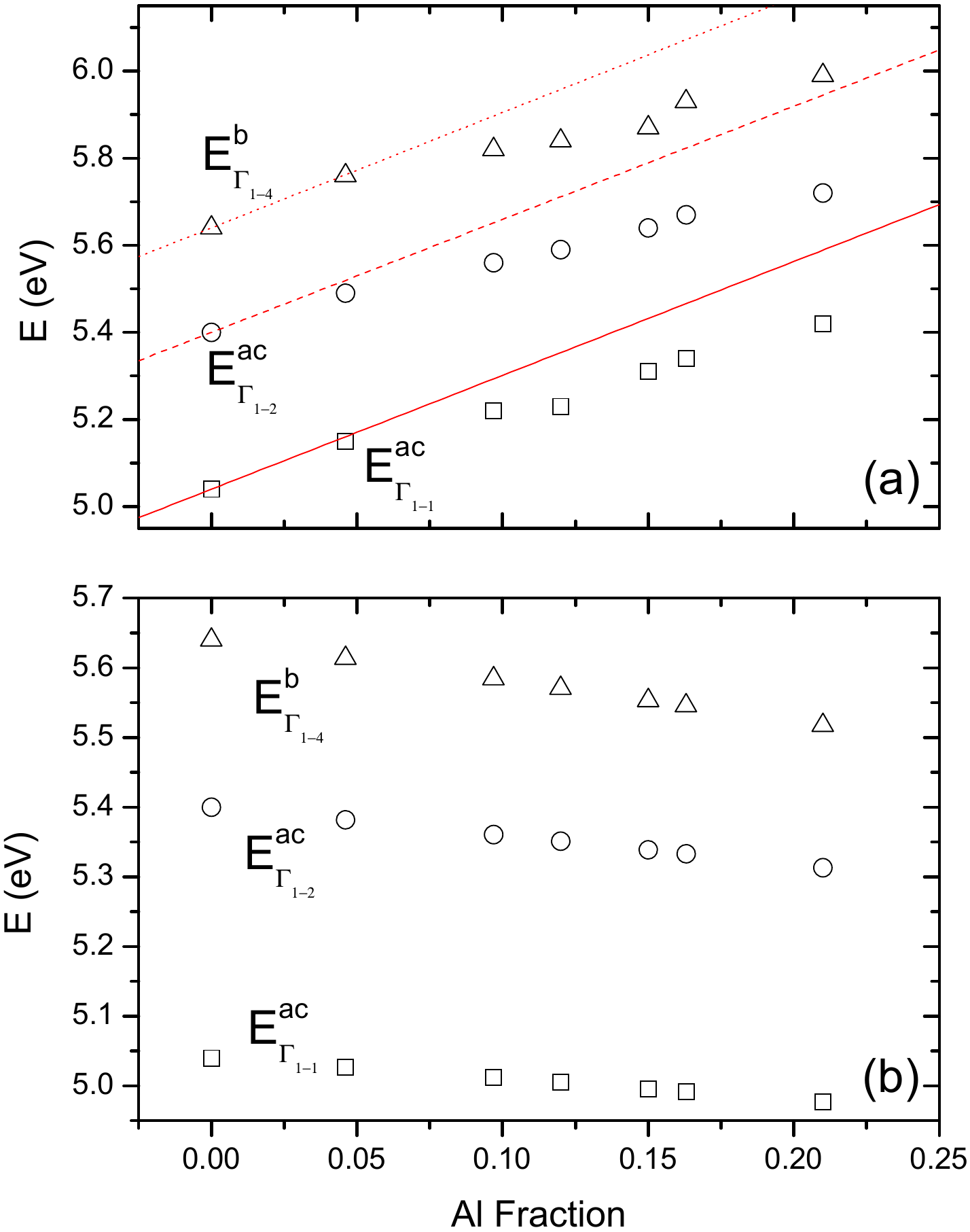}
    \caption{(a) Experimentally determined energy values of the lowest band-to-band transitions obtained from GSE. Lines depict the linear slopes of the respective transitions due to alloying predicted from DFT calculations. Different line styles pertain to different transitions. (b) DFT calculated effects of strain on the lowest band-to-band transitions according to Eqs.~\ref{eq:eg11}-\ref{eq:eg14} without the composition dependence of the strain free energy levels for the same compositions where experimental data are plotted in (a).}
    \label{fig:transitions}
  \end{center}
\end{figure}

The $\beta$-Ga$_2$O$_3$ substrates are monoclinic and their optical and spectroscopic properties were studied extensively, both experimentally and theoretically.\cite{Sturm_2015,Sturm_2016,Furthmuller_2016, SchubertPRB2016,Mock_2017Ga2O3,Ratnaparkhe_2017} The details of the generalized spectroscopic ellipsometry measurements and the analysis of the results follow the same model approach described by Mock~\textit{et al.},\cite{Mock_2017Ga2O3} which was applied earlier to a set of $\beta$-(Al$_{x}$Ga$_{1-x}$)$_2$O$_3$ thin films on (010) $\beta$-Ga$_2$O$_3$ substrates.\cite{Hilfiker_2019} Here, we augment the previous results with three additional samples. 

\subsubsection{Ellipsometry Measurements}

Generalized spectroscopic ellipsometry data was measured in ambient conditions in the Mueller matrix formalism\cite{Fujiwara_2007} from 1.5 to 9.0~eV using two separate instruments. Vacuum-ultra-violet (VUV) spectra was acquired from 6.5~eV to 9~eV using a commercially available rotating-analyzer ellipsometer (VUV-VASE, J.A. Woollam Co., Inc.). The rest of the spectral range was collected using a dual-rotating compensator ellipsometer (RC2, J.A. Woollam Co., Inc.), also commercial. All measurements were done at three angles of incidence ($\Phi_a$~=~50$^\circ$, 60$^\circ$, 70$^\circ$) and at multiple azimuthal angles. For VUV data, the sample was manually rotated in $\approx $~45$^\circ$ steps from 0$^\circ$ to 180$^\circ$ to fully capture the monoclinic symmetry. For NIR-Visible data (1.5~eV to 6.5~eV), the sample was rotated in 15$^\circ$ steps for a full rotation. Note that the VUV instrument does not permit acquisition of the 4th row of the Mueller matrix while the NIR-visible instrument does. This however does not hinder access to the full dielectric tensor in this case as $\beta$-(Al$_x$Ga$_{1-x}$)$_2$O$_3$ films are neither magnetically nor optically active. 

\subsubsection{Analysis of the dielectric tensor}
To analyze the ellipsometry data, best-match model calculations are employed. Specifically, the substrate-film(layer)-roughness(layer)-ambient model is used here.\cite{Fujiwara_2007} Figure~\ref{fig:Ga2O3unitcell} shows the coordinate system chosen here common to both the $\beta$-Ga$_2$O$_3$ substrate and  $\beta$-(Al$_x$Ga$_{1-x}$)$_2$O$_3$ film. Here $x$ is parallel to lattice vector \textbf{a}, $y$ is parallel to \textbf{c$^{\star}$}, where the monoclinic plane is defined by \textbf{a} and \textbf{c$^{\star}$}, and $z$ is parallel to \textbf{b}. This together describes the dielectric tensor directions. For further discussion about the choice of coordinate system, see Ref.~\onlinecite{Mock_2017Ga2O3}.

\begin{equation}\label{eq:eps}
\varepsilon = \begin{pmatrix} \varepsilon_{xx} & \varepsilon_{xy} & 0\\ \varepsilon_{xy} & \varepsilon_{yy} & 0\\ 0 & 0 & \varepsilon_{zz}\end{pmatrix}.
\end{equation}

The full description of the dielectric tensor of pure $\beta$-Ga$_2$O$_3$ modeled using an eigendielectric vector summation approach was published in Ref.~[\onlinecite{Mock_2017Ga2O3}]. The same approach and functions were then applied to the dielectric tensor of $\beta$-(Al$_x$Ga$_{1-x}$)$_2$O$_3$ for $x$~$\leq$~0.21 films similar to those studied here.\cite{Hilfiker_2019} We use a critical-point (CP) model dielectric function (MDF) approach, projected within tensor dyadics to perform our best-match model calculations detailed in Refs.~\onlinecite{Mock_2017Ga2O3,Hilfiker_2019}. We permit CP model parameters (amplitude, broadening, transition energy, and angular orientation) to freely vary during best-match parameter calculations. Due to little observed sensitivity, the exciton binding energy parameters for all $\beta$-(Al$_x$Ga$_{1-x}$)$_2$O$_3$ films were kept constant to the values obtained for $\beta$-Ga$_2$O$_3$.\cite{Mock_2017Ga2O3} We account for surface roughness by adding a thin model layer containing an isotropic average of all tensor elements weighted with 50$\%$ void. The best match ellipsometry model parameters for the $\beta$-(Al$_x$Ga$_{1-x}$)$_2$O$_3$ film thickness and roughness layer thickness are 135.(6)~nm and 0.(7)~nm for the 4.6$\%$ sample, 131.(2)~nm and 0.(9)~nm for the 9.7$\%$ sample, 82.(4)~nm and 1.(2)~nm for the 12$\%$ sample, 120.(6)~nm and 1.(8)~nm for the 15$\%$ sample, 128.(3)~nm and 0.(8)~nm for the 16.3$\%$ sample, and 62.(9)~nm and 1.(2)~nm for the 21$\%$ sample, respectively. The last digit which is determined within the 90$\%$ confidence interval is indicated with parentheses. The resulting energies for the three lowest band-to-band transitions are listed in Table~\ref{tab:strain_SE}.

\begin{table}
\centering
\caption{Transition energies of $\beta$-(Al$_{x}$Ga$_{1-x}$)$_2$O$_3$ thin films on (010) $\beta$-Ga$_2$O$_3$ obtained from the experimental GSE analysis (upper part); linear interpolation of transition energies between $\beta$-Ga$_2$O$_3$ and $\theta$-Al$_2$O$_3$ as described in the text (middle part); the estimated strain present in $\beta$-(Al$_{x}$Ga$_{1-x}$)$_2$O$_3$ thin films (lower part).}
\label{tab:strain_SE}
\begin{ruledtabular}
\begin{tabular}{{l}{c}{c}{c}{c}{c}{c}}
&\multicolumn{6}{c}{Aluminum Concentration}\\
\hline 
& 4.6\% & 9.7\% & 12\% & 15\% &  16.3\%&21\%\\
\hline 
\multicolumn{7}{c}{Transition Energies (eV)(GSE)}\\
\hline 
E$^{\mathrm{ac}}_{0}$ & 5.15 & 5.22 & 5.23 & 5.31 & 5.34 & 5.42 \\
E$^{\mathrm{ac}}_{1}$ & 5.49 & 5.56 & 5.59 & 5.64 & 5.67 & 5.72 \\
E$^{\mathrm{b}}_{0}$ & 5.76 & 5.82 & 5.84 & 5.87 & 5.93 & 5.99 \\
\hline
\multicolumn{7}{c}{Transition Energies (eV)(Linear Interpolation)}\\
\hline 
$E^{ac}_{\Gamma_{1-1}}$ & 5.16 & 5.30 & 5.36 & 5.43 & 5.47 & 5.59\\
$E^{ac}_{\Gamma_{1-2}}$ & 5.52 & 5.65 & 5.71 & 5.79 & 5.83 & 5.95\\
$E^{b}_{\Gamma_{1-4}}$ & 5.76 & 5.90 & 5.96 & 6.04 & 6.07 & 6.20\\
\hline
\multicolumn{7}{c}{Strain parameters}\\
\hline 
$\epsilon_{xx}$ & 0.0016 & 0.0033 & 0.0041 & 0.0052 & 0.0056 & 0.0073 \\
$\epsilon_{xy}$ & -0.0003 & -0.0007 & -0.0008 & -0.0010 & -0.0011 & -0.0014\\
$\epsilon_{yy}$ & 0.0015 & 0.0031 & 0.0039 & 0.0049 & 0.0053 & 0.0068\\
$\epsilon_{zz}$ & -0.0010 & -0.0020 & -0.0024 & -0.0031 & -0.0033 & -0.0043\\
\end{tabular}
\end{ruledtabular}
\end{table}

\section{Discussion}

\begin{figure}[ht]
\centering
\makebox[\linewidth][c]{\includegraphics[width=1.0\linewidth]{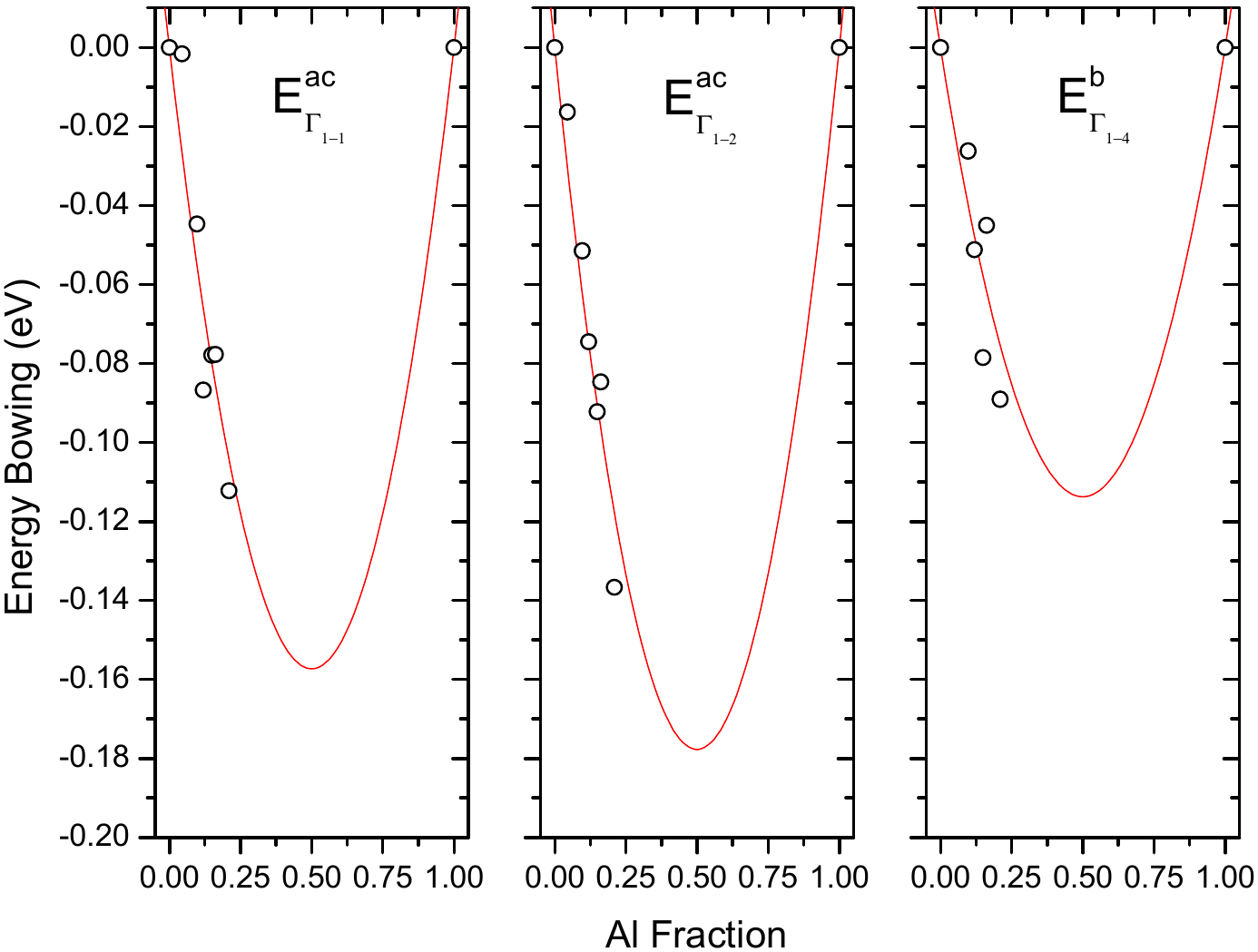}}
\centering
\caption{Experimentally (GSE) determined transition energy parameters reduced by (DFT) effects of strain and linear (Vegard's rule) composition shift (symbols) as shown in Figure~\ref{fig:transitions} and bowing $bx(x-1)$ dependencies (solid lines). The best-match bowing parameters are $b_{\Gamma_{1-1}}$ = 0.63~eV, $b_{\Gamma_{1-2}}$ = 0.71~eV, and $b_{\Gamma_{1-4}}$ = 0.45~eV.}
 \label{fig:Strain_Quad}
\end{figure}

The transition energies listed in Tab.~\ref{tab:strain_SE} and plotted in Figure~\ref{fig:transitions}, as obtained from the analysis of experimental GSE spectra, are a net result of multiple competing physical phenomena, predominantly alloying (the concentration of aluminum within the lattice of gallium oxide), and strain. We now have all the ingredients necessary to untangle these effects and extract the bowing dependence of transition energy versus the aluminum concentration. First, the linear portion of the concentration dependence can be obtained by linear interpolation of the band-to-band transition energies between the two end points corresponding to $\beta$-Ga$_2$O$_3$ ($x=0$) and $\theta$-Al$_2$O$_3$ ($x=1$). By comparing the transition energies obtained from DFT calculations. Here, we find linear band-to-band transition shifts between $\beta$-Ga$_2$O$_3$ and $\theta$-Al$_2$O$_3$ to be (Table~\ref{tab:slopes}) $\Gamma_{1-1}$:~2.613~eV, $\Gamma_{1-2}$:~2.594~eV, and $\Gamma_{1-4}$:~2.648~eV.

Then, the effects of strain can be accounted for by explicitly estimating the amount of strain in coherently grown films, as described above in Sec.~\ref{sec:samples}. The energy shifts caused by strain can then be calculated using equations~\ref{eq:eg11}-\ref{eq:eg14}. Note that we assume the equations to be valid, even though the actual amount of strain in some of the samples exceeds the $\pm 0.0035$ limit used in the determination of the deformation potentials. The values of strain for all samples studied are listed in Table~\ref{tab:strain_SE}. Note that strain values for $\epsilon_{xx},\epsilon_{yy},\epsilon_{xy}$ were calculated using the approach discussed in Sect.~\ref{sec:strainparameters}, while the $\epsilon_{zz}$ parameters were obtained from XRD measurements from the out-of-plane film lattice parameters. The resulting (strain-induced) energy shifts of the three band-to-band transitions are plotted in Fig.~\ref{fig:transitions}. After subtracting the effects of strain and the linear effect of alloying, the remaining effect in the transition energies can be attributed to the bandgap bowing. This is shown in Fig.~\ref{fig:Strain_Quad}. The dependence of the energy versus aluminum concentration is matched to $bx(x-1)$, and the bowing parameters are found to be $b_{\Gamma_{1-1}}$ = 0.63~eV, $b_{\Gamma_{1-2}}$ = 0.71~eV, $b_{\Gamma_{1-4}}$ = 0.45~eV, respectively, for the three lowest transitions. The resulting bowing parameters are of the same order of magnitude, but on the lower end of values reported in the literature so far, with the majority of results being reported for the indirect bandgap. 

To the best of our knowledge, no previous study considered individual Brillouin-zone center optical transitions, with an explicit treatment of excitonic effects, or accounted for the effects of strain in the alloy films. It is also worth mentioning that the outcome for the bowing parameters can be affected by the correction made for the linear composition dependencies of the bandgap energies between $\beta$-Ga$_2$O$_3$ and $\theta$-Al$_2$O$_3$. Especially the bandgap value of the latter is insecure, since $\theta$-Al$_2$O$_3$ does not exist as a bulk single crystal. Peelears~\textit{et al.} calculated values of 7.24~eV and 7.51~eV for the indirect and direct bandgaps of $\theta$-Al$_2$O$_3$, respectively,\cite{PeelaersAlGO} which were then assumed in the analysis by Jesenovec~\textit{et al.}\cite{Jesenovec_2022} Bhattacharjee~\textit{et al.}\cite{Bhattacharjee_2021} included the Al$_2$O$_3$ bandgap as a free parameter within their model, and the value obtained was ($6.8 \pm 0.2$)~eV. In our analysis, we used linear composition dependencies obtained from consistent DFT calculations, which we anchored to experimental transition energies for the $\beta$-Ga$_2$O$_3$ for a greater accuracy. 

Another interesting aspect worth mentioning here is the fact that equations~\ref{eq:eg11}-\ref{eq:eg14}, used to account for the strain shift of band-to-band transition energies, inherently include some amount of bowing. This is due to the fact that the strain components in these equations are also a linear function of composition, i.e., the samples with higher Aluminum concentration $x$ exhibit larger strain. This probably explains why among the values of bowing parameters gathered in Table~\ref{tab:bowing_lit}, the results not affected by strain (like those obtained from DFT calculations performed on unstrained equilibrium unit cells) tend to occupy the lower end of this broad range of values. If this is the case, it emphasizes the importance of strain for the properties of thin films, and ultimately in electronic devices.

\section{Conclusion}
The evolution of the three lowest direct band-to-band transition energies in $\beta$-(Al$_{x}$Ga$_{1-x}$)$_2$O$_3$ alloys with $x \le 0.21$ was studied using a combination of DFT calculations and GSE measurements. The DFT calculations provide a method for quantifying the role of coherent strain present in the $\beta$-(Al$_{x}$Ga$_{1-x}$)$_2$O$_3$ thin films on (010) $\beta$-Ga$_2$O$_3$ substrates. This allowed for an untangling of strain from the band-to-band transition energies determined using the spectroscopic ellipsometry eigenpolarization model approach. We report three individual bowing parameters associated with the three lowest band-to-band transitions in $\beta$-(Al$_{x}$Ga$_{1-x}$)$_2$O$_3$. The results described in this work are important for future design of electronic and optoelectronic device structures based on $\beta$-(Al$_{x}$Ga$_{1-x}$)$_2$O$_3$. 

\begin{acknowledgments}
This work was supported in part by the National Science Foundation (NSF) under awards NSF DMR 1808715 and NSF/EPSCoR RII Track-1: Emergent Quantum Materials and Technologies (EQUATE), Award OIA-2044049, by Air Force Office of Scientific Research under awards FA9550-18-1-0360, FA9550-19-S-0003, and FA9550-21-1-0259, by the Knut and Alice Wallenbergs Foundation award 'Wide-bandgap semiconductors for next generation quantum components', the Swedish Energy Agency under Award No. P45396-1, the Swedish Governmental Agency for Innovation Systems (VINNOVA) under the Competence Center Program Grant No. 2016-05190, the Swedish Research Council VR under Grands No. 2016-00889, Swedish Foundation for Strategic Research under Grants No. RIF14-055, and No. EM16-0024, and the Swedish Government
Strategic Research Area in Materials Science on Functional Materials at Link\"oping University, Faculty Grant SFO Mat LiU No. 2009-00971. M.S. acknowledges the University of Nebraska Foundation and the J. A. Woollam Foundation for financial support. J.S.S, A.M., and Y.Z. acknowledge funding from AFOSR through programs FA9550-18-1-0059 and FA9550-18-1-0479, and DTRA through program HDTRA 11710034.
\end{acknowledgments}


\begin{thebibliography}{45}%
\makeatletter
\providecommand \@ifxundefined [1]{%
 \@ifx{#1\undefined}
}%
\providecommand \@ifnum [1]{%
 \ifnum #1\expandafter \@firstoftwo
 \else \expandafter \@secondoftwo
 \fi
}%
\providecommand \@ifx [1]{%
 \ifx #1\expandafter \@firstoftwo
 \else \expandafter \@secondoftwo
 \fi
}%
\providecommand \natexlab [1]{#1}%
\providecommand \enquote  [1]{``#1''}%
\providecommand \bibnamefont  [1]{#1}%
\providecommand \bibfnamefont [1]{#1}%
\providecommand \citenamefont [1]{#1}%
\providecommand \href@noop [0]{\@secondoftwo}%
\providecommand \href [0]{\begingroup \@sanitize@url \@href}%
\providecommand \@href[1]{\@@startlink{#1}\@@href}%
\providecommand \@@href[1]{\endgroup#1\@@endlink}%
\providecommand \@sanitize@url [0]{\catcode `\\12\catcode `\$12\catcode
  `\&12\catcode `\#12\catcode `\^12\catcode `\_12\catcode `\%12\relax}%
\providecommand \@@startlink[1]{}%
\providecommand \@@endlink[0]{}%
\providecommand \url  [0]{\begingroup\@sanitize@url \@url }%
\providecommand \@url [1]{\endgroup\@href {#1}{\urlprefix }}%
\providecommand \urlprefix  [0]{URL }%
\providecommand \Eprint [0]{\href }%
\providecommand \doibase [0]{http://dx.doi.org/}%
\providecommand \selectlanguage [0]{\@gobble}%
\providecommand \bibinfo  [0]{\@secondoftwo}%
\providecommand \bibfield  [0]{\@secondoftwo}%
\providecommand \translation [1]{[#1]}%
\providecommand \BibitemOpen [0]{}%
\providecommand \bibitemStop [0]{}%
\providecommand \bibitemNoStop [0]{.\EOS\space}%
\providecommand \EOS [0]{\spacefactor3000\relax}%
\providecommand \BibitemShut  [1]{\csname bibitem#1\endcsname}%
\let\auto@bib@innerbib\@empty
\bibitem [{\citenamefont {Mock}\ \emph {et~al.}(2017)\citenamefont {Mock},
  \citenamefont {Korlacki}, \citenamefont {Briley}, \citenamefont
  {Darakchieva}, \citenamefont {Monemar}, \citenamefont {Kumagai},
  \citenamefont {Goto}, \citenamefont {Higashiwaki},\ and\ \citenamefont
  {Schubert}}]{Mock_2017Ga2O3}%
  \BibitemOpen
  \bibfield  {author} {\bibinfo {author} {\bibfnamefont {A.}~\bibnamefont
  {Mock}}, \bibinfo {author} {\bibfnamefont {R.}~\bibnamefont {Korlacki}},
  \bibinfo {author} {\bibfnamefont {C.}~\bibnamefont {Briley}}, \bibinfo
  {author} {\bibfnamefont {V.}~\bibnamefont {Darakchieva}}, \bibinfo {author}
  {\bibfnamefont {B.}~\bibnamefont {Monemar}}, \bibinfo {author} {\bibfnamefont
  {Y.}~\bibnamefont {Kumagai}}, \bibinfo {author} {\bibfnamefont
  {K.}~\bibnamefont {Goto}}, \bibinfo {author} {\bibfnamefont {M.}~\bibnamefont
  {Higashiwaki}}, \ and\ \bibinfo {author} {\bibfnamefont {M.}~\bibnamefont
  {Schubert}},\ }\href {\doibase 10.1103/PhysRevB.96.245205} {\bibfield
  {journal} {\bibinfo  {journal} {Phys.\ Rev.\ B}\ }\textbf {\bibinfo {volume}
  {96}},\ \bibinfo {pages} {245205} (\bibinfo {year} {2017})}\BibitemShut
  {NoStop}%
\bibitem [{\citenamefont {Harman}\ \emph {et~al.}(1994)\citenamefont {Harman},
  \citenamefont {Ninomiya},\ and\ \citenamefont
  {Adachi}}]{doi:10.1063/1.357922}%
  \BibitemOpen
  \bibfield  {author} {\bibinfo {author} {\bibfnamefont {A.~K.}\ \bibnamefont
  {Harman}}, \bibinfo {author} {\bibfnamefont {S.}~\bibnamefont {Ninomiya}}, \
  and\ \bibinfo {author} {\bibfnamefont {S.}~\bibnamefont {Adachi}},\ }\href
  {\doibase 10.1063/1.357922} {\bibfield  {journal} {\bibinfo  {journal} {J.
  Appl. Phys.}\ }\textbf {\bibinfo {volume} {76}},\ \bibinfo {pages} {8032}
  (\bibinfo {year} {1994})}\BibitemShut {NoStop}%
\bibitem [{\citenamefont {Slobodyan}\ \emph {et~al.}(2022)\citenamefont
  {Slobodyan}, \citenamefont {Flicker}, \citenamefont {Dickerson},
  \citenamefont {Shoemaker}, \citenamefont {Binder}, \citenamefont {Smith},
  \citenamefont {Goodnick}, \citenamefont {Kaplar},\ and\ \citenamefont
  {Hollis}}]{Slobodyan2022}%
  \BibitemOpen
  \bibfield  {author} {\bibinfo {author} {\bibfnamefont {O.}~\bibnamefont
  {Slobodyan}}, \bibinfo {author} {\bibfnamefont {J.}~\bibnamefont {Flicker}},
  \bibinfo {author} {\bibfnamefont {J.}~\bibnamefont {Dickerson}}, \bibinfo
  {author} {\bibfnamefont {J.}~\bibnamefont {Shoemaker}}, \bibinfo {author}
  {\bibfnamefont {A.}~\bibnamefont {Binder}}, \bibinfo {author} {\bibfnamefont
  {T.}~\bibnamefont {Smith}}, \bibinfo {author} {\bibfnamefont
  {S.}~\bibnamefont {Goodnick}}, \bibinfo {author} {\bibfnamefont
  {R.}~\bibnamefont {Kaplar}}, \ and\ \bibinfo {author} {\bibfnamefont
  {M.}~\bibnamefont {Hollis}},\ }\href {\doibase 10.1557/s43578-021-00465-2}
  {\bibfield  {journal} {\bibinfo  {journal} {Journal of Materials Research}\
  }\textbf {\bibinfo {volume} {37}},\ \bibinfo {pages} {849} (\bibinfo {year}
  {2022})}\BibitemShut {NoStop}%
\bibitem [{\citenamefont {Spencer}\ \emph {et~al.}(2022)\citenamefont
  {Spencer}, \citenamefont {Mock}, \citenamefont {Jacobs}, \citenamefont
  {Schubert}, \citenamefont {Zhang},\ and\ \citenamefont
  {Tadjer}}]{doi:10.1063/5.0078037}%
  \BibitemOpen
  \bibfield  {author} {\bibinfo {author} {\bibfnamefont {J.~A.}\ \bibnamefont
  {Spencer}}, \bibinfo {author} {\bibfnamefont {A.~L.}\ \bibnamefont {Mock}},
  \bibinfo {author} {\bibfnamefont {A.~G.}\ \bibnamefont {Jacobs}}, \bibinfo
  {author} {\bibfnamefont {M.}~\bibnamefont {Schubert}}, \bibinfo {author}
  {\bibfnamefont {Y.}~\bibnamefont {Zhang}}, \ and\ \bibinfo {author}
  {\bibfnamefont {M.~J.}\ \bibnamefont {Tadjer}},\ }\href {\doibase
  10.1063/5.0078037} {\bibfield  {journal} {\bibinfo  {journal} {Applied
  Physics Reviews}\ }\textbf {\bibinfo {volume} {9}},\ \bibinfo {pages}
  {011315} (\bibinfo {year} {2022})}\BibitemShut {NoStop}%
\bibitem [{\citenamefont {Schubert}\ \emph {et~al.}(2016)\citenamefont
  {Schubert}, \citenamefont {Korlacki}, \citenamefont {Knight}, \citenamefont
  {Hofmann}, \citenamefont {Sch\"{o}che}, \citenamefont {Darakchieva},
  \citenamefont {Janz\'{e}n}, \citenamefont {Monemar}, \citenamefont {Gogova},
  \citenamefont {Thieu}, \citenamefont {Togashi}, \citenamefont {Murakami},
  \citenamefont {Kumagai}, \citenamefont {Goto}, \citenamefont {Kuramata},
  \citenamefont {Yamakoshi},\ and\ \citenamefont
  {Higashiwaki}}]{SchubertPRB2016}%
  \BibitemOpen
  \bibfield  {author} {\bibinfo {author} {\bibfnamefont {M.}~\bibnamefont
  {Schubert}}, \bibinfo {author} {\bibfnamefont {R.}~\bibnamefont {Korlacki}},
  \bibinfo {author} {\bibfnamefont {S.}~\bibnamefont {Knight}}, \bibinfo
  {author} {\bibfnamefont {T.}~\bibnamefont {Hofmann}}, \bibinfo {author}
  {\bibfnamefont {S.}~\bibnamefont {Sch\"{o}che}}, \bibinfo {author}
  {\bibfnamefont {V.}~\bibnamefont {Darakchieva}}, \bibinfo {author}
  {\bibfnamefont {E.}~\bibnamefont {Janz\'{e}n}}, \bibinfo {author}
  {\bibfnamefont {B.}~\bibnamefont {Monemar}}, \bibinfo {author} {\bibfnamefont
  {D.}~\bibnamefont {Gogova}}, \bibinfo {author} {\bibfnamefont {Q.-T.}\
  \bibnamefont {Thieu}}, \bibinfo {author} {\bibfnamefont {R.}~\bibnamefont
  {Togashi}}, \bibinfo {author} {\bibfnamefont {H.}~\bibnamefont {Murakami}},
  \bibinfo {author} {\bibfnamefont {Y.}~\bibnamefont {Kumagai}}, \bibinfo
  {author} {\bibfnamefont {K.}~\bibnamefont {Goto}}, \bibinfo {author}
  {\bibfnamefont {A.}~\bibnamefont {Kuramata}}, \bibinfo {author}
  {\bibfnamefont {S.}~\bibnamefont {Yamakoshi}}, \ and\ \bibinfo {author}
  {\bibfnamefont {M.}~\bibnamefont {Higashiwaki}},\ }\href {\doibase
  10.1103/PhysRevB.93.125209} {\bibfield  {journal} {\bibinfo  {journal} {Phys.
  Rev. B}\ }\textbf {\bibinfo {volume} {93}},\ \bibinfo {pages} {125209}
  (\bibinfo {year} {2016})}\BibitemShut {NoStop}%
\bibitem [{\citenamefont {Peelaers}\ \emph {et~al.}(2018)\citenamefont
  {Peelaers}, \citenamefont {Varley}, \citenamefont {Speck},\ and\
  \citenamefont {Van~de Walle}}]{PeelaersAlGO}%
  \BibitemOpen
  \bibfield  {author} {\bibinfo {author} {\bibfnamefont {H.}~\bibnamefont
  {Peelaers}}, \bibinfo {author} {\bibfnamefont {J.~B.}\ \bibnamefont
  {Varley}}, \bibinfo {author} {\bibfnamefont {J.~S.}\ \bibnamefont {Speck}}, \
  and\ \bibinfo {author} {\bibfnamefont {C.~G.}\ \bibnamefont {Van~de Walle}},\
  }\href {\doibase 10.1063/1.5036991} {\bibfield  {journal} {\bibinfo
  {journal} {Appl.\ Phys.\ Lett.}\ }\textbf {\bibinfo {volume} {112}},\
  \bibinfo {pages} {242101} (\bibinfo {year} {2018})}\BibitemShut {NoStop}%
\bibitem [{\citenamefont {Shinohara}\ and\ \citenamefont
  {Fujita}(2008)}]{Shinohara_2008}%
  \BibitemOpen
  \bibfield  {author} {\bibinfo {author} {\bibfnamefont {D.}~\bibnamefont
  {Shinohara}}\ and\ \bibinfo {author} {\bibfnamefont {S.}~\bibnamefont
  {Fujita}},\ }\href {\doibase 10.1143/jjap.47.7311} {\bibfield  {journal}
  {\bibinfo  {journal} {Japanese Journal of Applied Physics}\ }\textbf
  {\bibinfo {volume} {47}},\ \bibinfo {pages} {7311} (\bibinfo {year}
  {2008})}\BibitemShut {NoStop}%
\bibitem [{\citenamefont {Jinno}\ \emph {et~al.}(2021)\citenamefont {Jinno},
  \citenamefont {Chang}, \citenamefont {Onuma}, \citenamefont {Cho},
  \citenamefont {Ho}, \citenamefont {Rowe}, \citenamefont {Cao}, \citenamefont
  {Lee}, \citenamefont {Protasenko}, \citenamefont {Schlom}, \citenamefont
  {Muller}, \citenamefont {Xing},\ and\ \citenamefont
  {Jena}}]{doi:10.1126/sciadv.abd5891}%
  \BibitemOpen
  \bibfield  {author} {\bibinfo {author} {\bibfnamefont {R.}~\bibnamefont
  {Jinno}}, \bibinfo {author} {\bibfnamefont {C.~S.}\ \bibnamefont {Chang}},
  \bibinfo {author} {\bibfnamefont {T.}~\bibnamefont {Onuma}}, \bibinfo
  {author} {\bibfnamefont {Y.}~\bibnamefont {Cho}}, \bibinfo {author}
  {\bibfnamefont {S.-T.}\ \bibnamefont {Ho}}, \bibinfo {author} {\bibfnamefont
  {D.}~\bibnamefont {Rowe}}, \bibinfo {author} {\bibfnamefont {M.~C.}\
  \bibnamefont {Cao}}, \bibinfo {author} {\bibfnamefont {K.}~\bibnamefont
  {Lee}}, \bibinfo {author} {\bibfnamefont {V.}~\bibnamefont {Protasenko}},
  \bibinfo {author} {\bibfnamefont {D.~G.}\ \bibnamefont {Schlom}}, \bibinfo
  {author} {\bibfnamefont {D.~A.}\ \bibnamefont {Muller}}, \bibinfo {author}
  {\bibfnamefont {H.~G.}\ \bibnamefont {Xing}}, \ and\ \bibinfo {author}
  {\bibfnamefont {D.}~\bibnamefont {Jena}},\ }\href {\doibase
  10.1126/sciadv.abd5891} {\bibfield  {journal} {\bibinfo  {journal} {Science
  Advances}\ }\textbf {\bibinfo {volume} {7}},\ \bibinfo {pages} {eabd5891}
  (\bibinfo {year} {2021})}\BibitemShut {NoStop}%
\bibitem [{\citenamefont {Wakabayashi}\ \emph {et~al.}(2021)\citenamefont
  {Wakabayashi}, \citenamefont {Yoshimatsu}, \citenamefont {Hattori},
  \citenamefont {Lee}, \citenamefont {Sakata},\ and\ \citenamefont
  {Ohtomo}}]{Wakabayashi_2021}%
  \BibitemOpen
  \bibfield  {author} {\bibinfo {author} {\bibfnamefont {R.}~\bibnamefont
  {Wakabayashi}}, \bibinfo {author} {\bibfnamefont {K.}~\bibnamefont
  {Yoshimatsu}}, \bibinfo {author} {\bibfnamefont {M.}~\bibnamefont {Hattori}},
  \bibinfo {author} {\bibfnamefont {J.-S.}\ \bibnamefont {Lee}}, \bibinfo
  {author} {\bibfnamefont {O.}~\bibnamefont {Sakata}}, \ and\ \bibinfo {author}
  {\bibfnamefont {A.}~\bibnamefont {Ohtomo}},\ }\href {\doibase
  10.1021/acs.cgd.1c00030} {\bibfield  {journal} {\bibinfo  {journal} {Cryst.\
  Growth.\ Des.}\ }\textbf {\bibinfo {volume} {21}},\ \bibinfo {pages} {2844}
  (\bibinfo {year} {2021})}\BibitemShut {NoStop}%
\bibitem [{\citenamefont {Hilfiker}\ \emph {et~al.}(2021)\citenamefont
  {Hilfiker}, \citenamefont {Korlacki}, \citenamefont {Jinno}, \citenamefont
  {Cho}, \citenamefont {Xing}, \citenamefont {Jena}, \citenamefont {Kilic},
  \citenamefont {Stokey},\ and\ \citenamefont
  {Schubert}}]{doi:10.1063/5.0031424}%
  \BibitemOpen
  \bibfield  {author} {\bibinfo {author} {\bibfnamefont {M.}~\bibnamefont
  {Hilfiker}}, \bibinfo {author} {\bibfnamefont {R.}~\bibnamefont {Korlacki}},
  \bibinfo {author} {\bibfnamefont {R.}~\bibnamefont {Jinno}}, \bibinfo
  {author} {\bibfnamefont {Y.}~\bibnamefont {Cho}}, \bibinfo {author}
  {\bibfnamefont {H.~G.}\ \bibnamefont {Xing}}, \bibinfo {author}
  {\bibfnamefont {D.}~\bibnamefont {Jena}}, \bibinfo {author} {\bibfnamefont
  {U.}~\bibnamefont {Kilic}}, \bibinfo {author} {\bibfnamefont
  {M.}~\bibnamefont {Stokey}}, \ and\ \bibinfo {author} {\bibfnamefont
  {M.}~\bibnamefont {Schubert}},\ }\href {\doibase 10.1063/5.0031424}
  {\bibfield  {journal} {\bibinfo  {journal} {Applied Physics Letters}\
  }\textbf {\bibinfo {volume} {118}},\ \bibinfo {pages} {062103} (\bibinfo
  {year} {2021})}\BibitemShut {NoStop}%
\bibitem [{\citenamefont {Sturm}\ \emph {et~al.}(2016)\citenamefont {Sturm},
  \citenamefont {Schmidt-Grund}, \citenamefont {Kranert}, \citenamefont
  {Furthm{\"u}ller}, \citenamefont {Bechstedt},\ and\ \citenamefont
  {Grundmann}}]{Sturm_2016}%
  \BibitemOpen
  \bibfield  {author} {\bibinfo {author} {\bibfnamefont {C.}~\bibnamefont
  {Sturm}}, \bibinfo {author} {\bibfnamefont {R.}~\bibnamefont
  {Schmidt-Grund}}, \bibinfo {author} {\bibfnamefont {C.}~\bibnamefont
  {Kranert}}, \bibinfo {author} {\bibfnamefont {J.}~\bibnamefont
  {Furthm{\"u}ller}}, \bibinfo {author} {\bibfnamefont {F.}~\bibnamefont
  {Bechstedt}}, \ and\ \bibinfo {author} {\bibfnamefont {M.}~\bibnamefont
  {Grundmann}},\ }\href {\doibase 10.1103/PhysRevB.94.035148} {\bibfield
  {journal} {\bibinfo  {journal} {Phys.\ Rev.\ B}\ }\textbf {\bibinfo {volume}
  {94}},\ \bibinfo {pages} {035148} (\bibinfo {year} {2016})}\BibitemShut
  {NoStop}%
\bibitem [{\citenamefont {Ratnaparkhe}\ and\ \citenamefont
  {Lambrecht}(2017)}]{Ratnaparkhe_2017}%
  \BibitemOpen
  \bibfield  {author} {\bibinfo {author} {\bibfnamefont {A.}~\bibnamefont
  {Ratnaparkhe}}\ and\ \bibinfo {author} {\bibfnamefont {W.~R.~L.}\
  \bibnamefont {Lambrecht}},\ }\href {\doibase 10.1063/1.4978668} {\bibfield
  {journal} {\bibinfo  {journal} {Appl.\ Phys.\ Lett.}\ }\textbf {\bibinfo
  {volume} {110}},\ \bibinfo {pages} {132103} (\bibinfo {year}
  {2017})}\BibitemShut {NoStop}%
\bibitem [{\citenamefont {Korlacki}\ \emph {et~al.}(2020)\citenamefont
  {Korlacki}, \citenamefont {Stokey}, \citenamefont {Mock}, \citenamefont
  {Knight}, \citenamefont {Papamichail}, \citenamefont {Darakchieva},\ and\
  \citenamefont {Schubert}}]{Korlacki2020}%
  \BibitemOpen
  \bibfield  {author} {\bibinfo {author} {\bibfnamefont {R.}~\bibnamefont
  {Korlacki}}, \bibinfo {author} {\bibfnamefont {M.}~\bibnamefont {Stokey}},
  \bibinfo {author} {\bibfnamefont {A.}~\bibnamefont {Mock}}, \bibinfo {author}
  {\bibfnamefont {S.}~\bibnamefont {Knight}}, \bibinfo {author} {\bibfnamefont
  {A.}~\bibnamefont {Papamichail}}, \bibinfo {author} {\bibfnamefont
  {V.}~\bibnamefont {Darakchieva}}, \ and\ \bibinfo {author} {\bibfnamefont
  {M.}~\bibnamefont {Schubert}},\ }\href {\doibase 10.1103/PhysRevB.102.180101}
  {\bibfield  {journal} {\bibinfo  {journal} {Phys. Rev. B}\ }\textbf {\bibinfo
  {volume} {102}},\ \bibinfo {pages} {180101(R)} (\bibinfo {year}
  {2020})}\BibitemShut {NoStop}%
\bibitem [{\citenamefont {Korlacki}\ \emph {et~al.}(2022)\citenamefont
  {Korlacki}, \citenamefont {Knudtson}, \citenamefont {Stokey}, \citenamefont
  {Hilfiker}, \citenamefont {Darakchieva},\ and\ \citenamefont
  {Schubert}}]{Korlacki_2022}%
  \BibitemOpen
  \bibfield  {author} {\bibinfo {author} {\bibfnamefont {R.}~\bibnamefont
  {Korlacki}}, \bibinfo {author} {\bibfnamefont {J.}~\bibnamefont {Knudtson}},
  \bibinfo {author} {\bibfnamefont {M.}~\bibnamefont {Stokey}}, \bibinfo
  {author} {\bibfnamefont {M.}~\bibnamefont {Hilfiker}}, \bibinfo {author}
  {\bibfnamefont {V.}~\bibnamefont {Darakchieva}}, \ and\ \bibinfo {author}
  {\bibfnamefont {M.}~\bibnamefont {Schubert}},\ }\href {\doibase
  10.1063/5.0078157} {\bibfield  {journal} {\bibinfo  {journal} {Appl.\ Phys.\
  Lett.}\ }\textbf {\bibinfo {volume} {120}},\ \bibinfo {pages} {042103}
  (\bibinfo {year} {2022})}\BibitemShut {NoStop}%
\bibitem [{\citenamefont {Magri}\ \emph {et~al.}(1991)\citenamefont {Magri},
  \citenamefont {Froyen},\ and\ \citenamefont {Zunger}}]{PhysRevB.44.7947}%
  \BibitemOpen
  \bibfield  {author} {\bibinfo {author} {\bibfnamefont {R.}~\bibnamefont
  {Magri}}, \bibinfo {author} {\bibfnamefont {S.}~\bibnamefont {Froyen}}, \
  and\ \bibinfo {author} {\bibfnamefont {A.}~\bibnamefont {Zunger}},\ }\href
  {\doibase 10.1103/PhysRevB.44.7947} {\bibfield  {journal} {\bibinfo
  {journal} {Phys. Rev. B}\ }\textbf {\bibinfo {volume} {44}},\ \bibinfo
  {pages} {7947} (\bibinfo {year} {1991})}\BibitemShut {NoStop}%
\bibitem [{\citenamefont {Wei}\ and\ \citenamefont
  {Zunger}(1994)}]{PhysRevB.49.14337}%
  \BibitemOpen
  \bibfield  {author} {\bibinfo {author} {\bibfnamefont {S.-H.}\ \bibnamefont
  {Wei}}\ and\ \bibinfo {author} {\bibfnamefont {A.}~\bibnamefont {Zunger}},\
  }\href {\doibase 10.1103/PhysRevB.49.14337} {\bibfield  {journal} {\bibinfo
  {journal} {Phys. Rev. B}\ }\textbf {\bibinfo {volume} {49}},\ \bibinfo
  {pages} {14337} (\bibinfo {year} {1994})}\BibitemShut {NoStop}%
\bibitem [{\citenamefont {Wei}\ \emph {et~al.}(2000)\citenamefont {Wei},
  \citenamefont {Zhang},\ and\ \citenamefont {Zunger}}]{doi:10.1063/1.372014}%
  \BibitemOpen
  \bibfield  {author} {\bibinfo {author} {\bibfnamefont {S.-H.}\ \bibnamefont
  {Wei}}, \bibinfo {author} {\bibfnamefont {S.~B.}\ \bibnamefont {Zhang}}, \
  and\ \bibinfo {author} {\bibfnamefont {A.}~\bibnamefont {Zunger}},\ }\href
  {\doibase 10.1063/1.372014} {\bibfield  {journal} {\bibinfo  {journal}
  {Journal of Applied Physics}\ }\textbf {\bibinfo {volume} {87}},\ \bibinfo
  {pages} {1304} (\bibinfo {year} {2000})}\BibitemShut {NoStop}%
\bibitem [{\citenamefont {Hilfiker}\ \emph {et~al.}(2022)\citenamefont
  {Hilfiker}, \citenamefont {Kilic}, \citenamefont {Stokey}, \citenamefont
  {Jinno}, \citenamefont {Cho}, \citenamefont {Xing}, \citenamefont {Jena},
  \citenamefont {Korlacki},\ and\ \citenamefont
  {Schubert}}]{HilfikeraAGOEg2022}%
  \BibitemOpen
  \bibfield  {author} {\bibinfo {author} {\bibfnamefont {M.}~\bibnamefont
  {Hilfiker}}, \bibinfo {author} {\bibfnamefont {U.}~\bibnamefont {Kilic}},
  \bibinfo {author} {\bibfnamefont {M.}~\bibnamefont {Stokey}}, \bibinfo
  {author} {\bibfnamefont {R.}~\bibnamefont {Jinno}}, \bibinfo {author}
  {\bibfnamefont {Y.}~\bibnamefont {Cho}}, \bibinfo {author} {\bibfnamefont
  {H.~G.}\ \bibnamefont {Xing}}, \bibinfo {author} {\bibfnamefont
  {D.}~\bibnamefont {Jena}}, \bibinfo {author} {\bibfnamefont {R.}~\bibnamefont
  {Korlacki}}, \ and\ \bibinfo {author} {\bibfnamefont {M.}~\bibnamefont
  {Schubert}},\ }\href {\doibase 10.1063/5.0087602} {\bibfield  {journal}
  {\bibinfo  {journal} {Applied Physics Letters}\ }\textbf {\bibinfo {volume}
  {121}},\ \bibinfo {pages} {052101} (\bibinfo {year} {2022})}\BibitemShut
  {NoStop}%
\bibitem [{\citenamefont {Hilfiker}\ \emph {et~al.}(2019)\citenamefont
  {Hilfiker}, \citenamefont {Kilic}, \citenamefont {Mock}, \citenamefont
  {Darakchieva}, \citenamefont {Knight}, \citenamefont {Korlacki},
  \citenamefont {Mauze}, \citenamefont {Zhang}, \citenamefont {Speck},\ and\
  \citenamefont {Schubert}}]{Hilfiker_2019}%
  \BibitemOpen
  \bibfield  {author} {\bibinfo {author} {\bibfnamefont {M.}~\bibnamefont
  {Hilfiker}}, \bibinfo {author} {\bibfnamefont {U.}~\bibnamefont {Kilic}},
  \bibinfo {author} {\bibfnamefont {A.}~\bibnamefont {Mock}}, \bibinfo {author}
  {\bibfnamefont {V.}~\bibnamefont {Darakchieva}}, \bibinfo {author}
  {\bibfnamefont {S.}~\bibnamefont {Knight}}, \bibinfo {author} {\bibfnamefont
  {R.}~\bibnamefont {Korlacki}}, \bibinfo {author} {\bibfnamefont
  {A.}~\bibnamefont {Mauze}}, \bibinfo {author} {\bibfnamefont
  {Y.}~\bibnamefont {Zhang}}, \bibinfo {author} {\bibfnamefont
  {J.}~\bibnamefont {Speck}}, \ and\ \bibinfo {author} {\bibfnamefont
  {M.}~\bibnamefont {Schubert}},\ }\href {\doibase 10.1063/1.5097780}
  {\bibfield  {journal} {\bibinfo  {journal} {Appl. Phys. Lett.}\ }\textbf
  {\bibinfo {volume} {114}},\ \bibinfo {pages} {231901} (\bibinfo {year}
  {2019})}\BibitemShut {NoStop}%
\bibitem [{\citenamefont {Kim}\ \emph {et~al.}(2021)\citenamefont {Kim},
  \citenamefont {Ko}, \citenamefont {Chung},\ and\ \citenamefont
  {Cho}}]{Kim_2021}%
  \BibitemOpen
  \bibfield  {author} {\bibinfo {author} {\bibfnamefont {H.~W.}\ \bibnamefont
  {Kim}}, \bibinfo {author} {\bibfnamefont {H.}~\bibnamefont {Ko}}, \bibinfo
  {author} {\bibfnamefont {Y.-C.}\ \bibnamefont {Chung}}, \ and\ \bibinfo
  {author} {\bibfnamefont {S.~B.}\ \bibnamefont {Cho}},\ }\href {\doibase
  10.1016/j.jeurceramsoc.2020.08.067} {\bibfield  {journal} {\bibinfo
  {journal} {J.\ Eur.\ Ceramic\ Soc.}\ }\textbf {\bibinfo {volume} {41}},\
  \bibinfo {pages} {611} (\bibinfo {year} {2021})}\BibitemShut {NoStop}%
\bibitem [{\citenamefont {Bhattacharjee}\ \emph {et~al.}(2021)\citenamefont
  {Bhattacharjee}, \citenamefont {Ghosh}, \citenamefont {Pokhriyal},
  \citenamefont {Gangwar}, \citenamefont {Dutt}, \citenamefont {Sagdeo},
  \citenamefont {Tiwari},\ and\ \citenamefont {Singh}}]{Bhattacharjee_2021}%
  \BibitemOpen
  \bibfield  {author} {\bibinfo {author} {\bibfnamefont {J.}~\bibnamefont
  {Bhattacharjee}}, \bibinfo {author} {\bibfnamefont {S.}~\bibnamefont
  {Ghosh}}, \bibinfo {author} {\bibfnamefont {P.}~\bibnamefont {Pokhriyal}},
  \bibinfo {author} {\bibfnamefont {R.}~\bibnamefont {Gangwar}}, \bibinfo
  {author} {\bibfnamefont {R.}~\bibnamefont {Dutt}}, \bibinfo {author}
  {\bibfnamefont {A.}~\bibnamefont {Sagdeo}}, \bibinfo {author} {\bibfnamefont
  {P.}~\bibnamefont {Tiwari}}, \ and\ \bibinfo {author} {\bibfnamefont {S.~D.}\
  \bibnamefont {Singh}},\ }\href {\doibase 10.1063/5.0055874} {\bibfield
  {journal} {\bibinfo  {journal} {AIP\ Advances}\ }\textbf {\bibinfo {volume}
  {11}},\ \bibinfo {pages} {075025} (\bibinfo {year} {2021})}\BibitemShut
  {NoStop}%
\bibitem [{\citenamefont {Wang}\ \emph {et~al.}(2018)\citenamefont {Wang},
  \citenamefont {Li}, \citenamefont {Ni},\ and\ \citenamefont
  {Janotti}}]{WangAlGODFT}%
  \BibitemOpen
  \bibfield  {author} {\bibinfo {author} {\bibfnamefont {T.}~\bibnamefont
  {Wang}}, \bibinfo {author} {\bibfnamefont {W.}~\bibnamefont {Li}}, \bibinfo
  {author} {\bibfnamefont {C.}~\bibnamefont {Ni}}, \ and\ \bibinfo {author}
  {\bibfnamefont {A.}~\bibnamefont {Janotti}},\ }\href {\doibase
  10.1103/PhysRevApplied.10.011003} {\bibfield  {journal} {\bibinfo  {journal}
  {Phys. Rev. Appl.}\ }\textbf {\bibinfo {volume} {10}},\ \bibinfo {pages}
  {011003} (\bibinfo {year} {2018})}\BibitemShut {NoStop}%
\bibitem [{\citenamefont {Ratnaparkhe}\ and\ \citenamefont
  {Lambrecht}(2020)}]{Ratnaparkhe_2020}%
  \BibitemOpen
  \bibfield  {author} {\bibinfo {author} {\bibfnamefont {A.}~\bibnamefont
  {Ratnaparkhe}}\ and\ \bibinfo {author} {\bibfnamefont {W.~R.~L.}\
  \bibnamefont {Lambrecht}},\ }\href {\doibase 10.1002/pssb.201900317}
  {\bibfield  {journal} {\bibinfo  {journal} {Phys.\ Status\ Solidi\ B}\
  }\textbf {\bibinfo {volume} {257}},\ \bibinfo {pages} {1900317} (\bibinfo
  {year} {2020})}\BibitemShut {NoStop}%
\bibitem [{\citenamefont {Bhuiyan}\ \emph {et~al.}(2020)\citenamefont
  {Bhuiyan}, \citenamefont {Feng}, \citenamefont {Johnson}, \citenamefont
  {Huang}, \citenamefont {Hwang},\ and\ \citenamefont {Zhao}}]{Bhuiyan_2020}%
  \BibitemOpen
  \bibfield  {author} {\bibinfo {author} {\bibfnamefont {A.~F. M. A.~U.}\
  \bibnamefont {Bhuiyan}}, \bibinfo {author} {\bibfnamefont {Z.}~\bibnamefont
  {Feng}}, \bibinfo {author} {\bibfnamefont {J.~M.}\ \bibnamefont {Johnson}},
  \bibinfo {author} {\bibfnamefont {H.-L.}\ \bibnamefont {Huang}}, \bibinfo
  {author} {\bibfnamefont {J.}~\bibnamefont {Hwang}}, \ and\ \bibinfo {author}
  {\bibfnamefont {H.}~\bibnamefont {Zhao}},\ }\href {\doibase
  10.1063/5.0031584} {\bibfield  {journal} {\bibinfo  {journal} {Appl.\ Phys.\
  Lett.}\ }\textbf {\bibinfo {volume} {117}},\ \bibinfo {pages} {252105}
  (\bibinfo {year} {2020})}\BibitemShut {NoStop}%
\bibitem [{\citenamefont {Li}\ \emph {et~al.}(2018)\citenamefont {Li},
  \citenamefont {Chen}, \citenamefont {Ma}, \citenamefont {Cui}, \citenamefont
  {Ren}, \citenamefont {Gu}, \citenamefont {Zhang}, \citenamefont {Zheng},
  \citenamefont {Ringer}, \citenamefont {Fu}, \citenamefont {Tan},
  \citenamefont {Jagadish},\ and\ \citenamefont {Ye}}]{Li_2018}%
  \BibitemOpen
  \bibfield  {author} {\bibinfo {author} {\bibfnamefont {J.}~\bibnamefont
  {Li}}, \bibinfo {author} {\bibfnamefont {X.}~\bibnamefont {Chen}}, \bibinfo
  {author} {\bibfnamefont {T.}~\bibnamefont {Ma}}, \bibinfo {author}
  {\bibfnamefont {X.}~\bibnamefont {Cui}}, \bibinfo {author} {\bibfnamefont
  {F.-F.}\ \bibnamefont {Ren}}, \bibinfo {author} {\bibfnamefont
  {S.}~\bibnamefont {Gu}}, \bibinfo {author} {\bibfnamefont {R.}~\bibnamefont
  {Zhang}}, \bibinfo {author} {\bibfnamefont {Y.}~\bibnamefont {Zheng}},
  \bibinfo {author} {\bibfnamefont {S.~P.}\ \bibnamefont {Ringer}}, \bibinfo
  {author} {\bibfnamefont {L.}~\bibnamefont {Fu}}, \bibinfo {author}
  {\bibfnamefont {H.~H.}\ \bibnamefont {Tan}}, \bibinfo {author} {\bibfnamefont
  {C.}~\bibnamefont {Jagadish}}, \ and\ \bibinfo {author} {\bibfnamefont
  {J.}~\bibnamefont {Ye}},\ }\href {\doibase 10.1063/1.5027763} {\bibfield
  {journal} {\bibinfo  {journal} {Appl.\ Phys.\ Lett.}\ }\textbf {\bibinfo
  {volume} {113}},\ \bibinfo {pages} {041901} (\bibinfo {year}
  {2018})}\BibitemShut {NoStop}%
\bibitem [{\citenamefont {Ota}(2020)}]{Ota_2020}%
  \BibitemOpen
  \bibfield  {author} {\bibinfo {author} {\bibfnamefont {Y.}~\bibnamefont
  {Ota}},\ }\href {\doibase 10.1063/5.00312241} {\bibfield  {journal} {\bibinfo
   {journal} {AIP\ Advances}\ }\textbf {\bibinfo {volume} {10}},\ \bibinfo
  {pages} {125321} (\bibinfo {year} {2020})}\BibitemShut {NoStop}%
\bibitem [{\citenamefont {Jesenovec}\ \emph {et~al.}(2022)\citenamefont
  {Jesenovec}, \citenamefont {Dutton}, \citenamefont {Stone-Weiss},
  \citenamefont {Chmielewski}, \citenamefont {Saleh}, \citenamefont {Peterson},
  \citenamefont {Alem}, \citenamefont {Krishnamoorthy},\ and\ \citenamefont
  {McCloy}}]{Jesenovec_2022}%
  \BibitemOpen
  \bibfield  {author} {\bibinfo {author} {\bibfnamefont {J.}~\bibnamefont
  {Jesenovec}}, \bibinfo {author} {\bibfnamefont {B.}~\bibnamefont {Dutton}},
  \bibinfo {author} {\bibfnamefont {N.}~\bibnamefont {Stone-Weiss}}, \bibinfo
  {author} {\bibfnamefont {A.}~\bibnamefont {Chmielewski}}, \bibinfo {author}
  {\bibfnamefont {M.}~\bibnamefont {Saleh}}, \bibinfo {author} {\bibfnamefont
  {C.}~\bibnamefont {Peterson}}, \bibinfo {author} {\bibfnamefont
  {N.}~\bibnamefont {Alem}}, \bibinfo {author} {\bibfnamefont {S.}~\bibnamefont
  {Krishnamoorthy}}, \ and\ \bibinfo {author} {\bibfnamefont {J.~S.}\
  \bibnamefont {McCloy}},\ }\href {\doibase 10.1063/5.0073502} {\bibfield
  {journal} {\bibinfo  {journal} {J.\ Appl.\ Phys.}\ }\textbf {\bibinfo
  {volume} {131}},\ \bibinfo {pages} {155702} (\bibinfo {year}
  {2022})}\BibitemShut {NoStop}%
\bibitem [{\citenamefont {Kranert}\ \emph {et~al.}(2015)\citenamefont
  {Kranert}, \citenamefont {Jenderka}, \citenamefont {Lenzner}, \citenamefont
  {Lorenz}, \citenamefont {von Wenckstern}, \citenamefont {Schmidt-Grund},\
  and\ \citenamefont {Grundmann}}]{Kranert_2015}%
  \BibitemOpen
  \bibfield  {author} {\bibinfo {author} {\bibfnamefont {C.}~\bibnamefont
  {Kranert}}, \bibinfo {author} {\bibfnamefont {M.}~\bibnamefont {Jenderka}},
  \bibinfo {author} {\bibfnamefont {J.}~\bibnamefont {Lenzner}}, \bibinfo
  {author} {\bibfnamefont {M.}~\bibnamefont {Lorenz}}, \bibinfo {author}
  {\bibfnamefont {H.}~\bibnamefont {von Wenckstern}}, \bibinfo {author}
  {\bibfnamefont {R.}~\bibnamefont {Schmidt-Grund}}, \ and\ \bibinfo {author}
  {\bibfnamefont {M.}~\bibnamefont {Grundmann}},\ }\href {\doibase
  10.1063/1.4915627} {\bibfield  {journal} {\bibinfo  {journal} {J.\ Appl.\
  Phys.}\ }\textbf {\bibinfo {volume} {117}},\ \bibinfo {pages} {125703}
  (\bibinfo {year} {2015})}\BibitemShut {NoStop}%
\bibitem [{\citenamefont {Oshima}\ \emph {et~al.}(2016)\citenamefont {Oshima},
  \citenamefont {Ahmadi}, \citenamefont {Badescu}, \citenamefont {Wu},\ and\
  \citenamefont {Speck}}]{OshimaAPE2016AGO}%
  \BibitemOpen
  \bibfield  {author} {\bibinfo {author} {\bibfnamefont {Y.}~\bibnamefont
  {Oshima}}, \bibinfo {author} {\bibfnamefont {E.}~\bibnamefont {Ahmadi}},
  \bibinfo {author} {\bibfnamefont {S.~C.}\ \bibnamefont {Badescu}}, \bibinfo
  {author} {\bibfnamefont {F.}~\bibnamefont {Wu}}, \ and\ \bibinfo {author}
  {\bibfnamefont {J.}~\bibnamefont {Speck}},\ }\href {\doibase
  10.7567/APEX.9.061102} {\bibfield  {journal} {\bibinfo  {journal} {Appl.\
  Phys.\ Exp.}\ }\textbf {\bibinfo {volume} {9}},\ \bibinfo {pages} {061102}
  (\bibinfo {year} {2016})}\BibitemShut {NoStop}%
\bibitem [{\citenamefont {Monkhorst}\ and\ \citenamefont
  {Pack}(1976)}]{Monkhorst1976}%
  \BibitemOpen
  \bibfield  {author} {\bibinfo {author} {\bibfnamefont {H.~J.}\ \bibnamefont
  {Monkhorst}}\ and\ \bibinfo {author} {\bibfnamefont {J.~D.}\ \bibnamefont
  {Pack}},\ }\href {\doibase 10.1103/PhysRevB.13.5188} {\bibfield  {journal}
  {\bibinfo  {journal} {Phys. Rev. B}\ }\textbf {\bibinfo {volume} {13}},\
  \bibinfo {pages} {5188} (\bibinfo {year} {1976})}\BibitemShut {NoStop}%
\bibitem [{\citenamefont {Giannozzi}\ \emph {et~al.}(2009)\citenamefont
  {Giannozzi}, \citenamefont {Baroni}, \citenamefont {Bonini}, \citenamefont
  {Calandra}, \citenamefont {Car}, \citenamefont {Cavazzoni}, \citenamefont
  {Ceresoli}, \citenamefont {Chiarotti}, \citenamefont {Cococcioni},
  \citenamefont {Dabo}, \citenamefont {Corso}, \citenamefont {de~Gironcoli},
  \citenamefont {Fabris}, \citenamefont {Fratesi}, \citenamefont {Gebauer},
  \citenamefont {Gerstmann}, \citenamefont {Gougoussis}, \citenamefont
  {Kokalj}, \citenamefont {Lazzeri}, \citenamefont {Martin-Samos},
  \citenamefont {Marzari}, \citenamefont {Mauri}, \citenamefont {Mazzarello},
  \citenamefont {Paolini}, \citenamefont {Pasquarello}, \citenamefont
  {Paulatto}, \citenamefont {Sbraccia}, \citenamefont {Scandolo}, \citenamefont
  {Sclauzero}, \citenamefont {Seitsonen}, \citenamefont {Smogunov},
  \citenamefont {Umari},\ and\ \citenamefont
  {Wentzcovitch}}]{GiannozziJPCM2009QE}%
  \BibitemOpen
  \bibfield  {author} {\bibinfo {author} {\bibfnamefont {P.}~\bibnamefont
  {Giannozzi}}, \bibinfo {author} {\bibfnamefont {S.}~\bibnamefont {Baroni}},
  \bibinfo {author} {\bibfnamefont {N.}~\bibnamefont {Bonini}}, \bibinfo
  {author} {\bibfnamefont {M.}~\bibnamefont {Calandra}}, \bibinfo {author}
  {\bibfnamefont {R.}~\bibnamefont {Car}}, \bibinfo {author} {\bibfnamefont
  {C.}~\bibnamefont {Cavazzoni}}, \bibinfo {author} {\bibfnamefont
  {D.}~\bibnamefont {Ceresoli}}, \bibinfo {author} {\bibfnamefont {G.~L.}\
  \bibnamefont {Chiarotti}}, \bibinfo {author} {\bibfnamefont {M.}~\bibnamefont
  {Cococcioni}}, \bibinfo {author} {\bibfnamefont {I.}~\bibnamefont {Dabo}},
  \bibinfo {author} {\bibfnamefont {A.~D.}\ \bibnamefont {Corso}}, \bibinfo
  {author} {\bibfnamefont {S.}~\bibnamefont {de~Gironcoli}}, \bibinfo {author}
  {\bibfnamefont {S.}~\bibnamefont {Fabris}}, \bibinfo {author} {\bibfnamefont
  {G.}~\bibnamefont {Fratesi}}, \bibinfo {author} {\bibfnamefont
  {R.}~\bibnamefont {Gebauer}}, \bibinfo {author} {\bibfnamefont
  {U.}~\bibnamefont {Gerstmann}}, \bibinfo {author} {\bibfnamefont
  {C.}~\bibnamefont {Gougoussis}}, \bibinfo {author} {\bibfnamefont
  {A.}~\bibnamefont {Kokalj}}, \bibinfo {author} {\bibfnamefont
  {M.}~\bibnamefont {Lazzeri}}, \bibinfo {author} {\bibfnamefont
  {L.}~\bibnamefont {Martin-Samos}}, \bibinfo {author} {\bibfnamefont
  {N.}~\bibnamefont {Marzari}}, \bibinfo {author} {\bibfnamefont
  {F.}~\bibnamefont {Mauri}}, \bibinfo {author} {\bibfnamefont
  {R.}~\bibnamefont {Mazzarello}}, \bibinfo {author} {\bibfnamefont
  {S.}~\bibnamefont {Paolini}}, \bibinfo {author} {\bibfnamefont
  {A.}~\bibnamefont {Pasquarello}}, \bibinfo {author} {\bibfnamefont
  {L.}~\bibnamefont {Paulatto}}, \bibinfo {author} {\bibfnamefont
  {C.}~\bibnamefont {Sbraccia}}, \bibinfo {author} {\bibfnamefont
  {S.}~\bibnamefont {Scandolo}}, \bibinfo {author} {\bibfnamefont
  {G.}~\bibnamefont {Sclauzero}}, \bibinfo {author} {\bibfnamefont {A.~P.}\
  \bibnamefont {Seitsonen}}, \bibinfo {author} {\bibfnamefont {A.}~\bibnamefont
  {Smogunov}}, \bibinfo {author} {\bibfnamefont {P.}~\bibnamefont {Umari}}, \
  and\ \bibinfo {author} {\bibfnamefont {R.~M.}\ \bibnamefont {Wentzcovitch}},\
  }\href {\doibase 10.1088/0953-8984/21/39/395502} {\bibfield  {journal}
  {\bibinfo  {journal} {J. Phys.: Cond. Mat.}\ }\textbf {\bibinfo {volume}
  {21}},\ \bibinfo {pages} {395502} (\bibinfo {year} {2009})}\BibitemShut
  {NoStop}%
\bibitem [{\citenamefont {Perdew}\ \emph {et~al.}(1996)\citenamefont {Perdew},
  \citenamefont {Burke},\ and\ \citenamefont {Ernzerhof}}]{PBE1996}%
  \BibitemOpen
  \bibfield  {author} {\bibinfo {author} {\bibfnamefont {J.~P.}\ \bibnamefont
  {Perdew}}, \bibinfo {author} {\bibfnamefont {K.}~\bibnamefont {Burke}}, \
  and\ \bibinfo {author} {\bibfnamefont {M.}~\bibnamefont {Ernzerhof}},\ }\href
  {\doibase 10.1103/PhysRevLett.77.3865} {\bibfield  {journal} {\bibinfo
  {journal} {Phys. Rev. Lett.}\ }\textbf {\bibinfo {volume} {77}},\ \bibinfo
  {pages} {3865} (\bibinfo {year} {1996})}\BibitemShut {NoStop}%
\bibitem [{\citenamefont {Fuchs}\ and\ \citenamefont
  {Scheffler}(1999)}]{FuchsCPC1999}%
  \BibitemOpen
  \bibfield  {author} {\bibinfo {author} {\bibfnamefont {M.}~\bibnamefont
  {Fuchs}}\ and\ \bibinfo {author} {\bibfnamefont {M.}~\bibnamefont
  {Scheffler}},\ }\href {\doibase 10.1016/S0010-4655(98)00201-X} {\bibfield
  {journal} {\bibinfo  {journal} {Comput. Phys. Commun.}\ }\textbf {\bibinfo
  {volume} {119}},\ \bibinfo {pages} {67} (\bibinfo {year} {1999})}\BibitemShut
  {NoStop}%
\bibitem [{\citenamefont {Troullier}\ and\ \citenamefont
  {Martins}(1991)}]{TroullierPRB1991}%
  \BibitemOpen
  \bibfield  {author} {\bibinfo {author} {\bibfnamefont {N.}~\bibnamefont
  {Troullier}}\ and\ \bibinfo {author} {\bibfnamefont {J.~L.}\ \bibnamefont
  {Martins}},\ }\href {\doibase 10.1103/PhysRevB.43.1993} {\bibfield  {journal}
  {\bibinfo  {journal} {Phys. Rev. B}\ }\textbf {\bibinfo {volume} {43}},\
  \bibinfo {pages} {1993} (\bibinfo {year} {1991})}\BibitemShut {NoStop}%
\bibitem [{\citenamefont {Setyawan}\ and\ \citenamefont
  {Curtarolo}(2010)}]{setyawan2010}%
  \BibitemOpen
  \bibfield  {author} {\bibinfo {author} {\bibfnamefont {W.}~\bibnamefont
  {Setyawan}}\ and\ \bibinfo {author} {\bibfnamefont {S.}~\bibnamefont
  {Curtarolo}},\ }\href {\doibase 10.1016/j.commatsci.2010.05.010} {\bibfield
  {journal} {\bibinfo  {journal} {Comput.\ Mat.\ Sci.}\ }\textbf {\bibinfo
  {volume} {49}},\ \bibinfo {pages} {299} (\bibinfo {year} {2010})}\BibitemShut
  {NoStop}%
\bibitem [{\citenamefont {Song}\ \emph {et~al.}(2011)\citenamefont {Song},
  \citenamefont {Yamashita},\ and\ \citenamefont {Hirao}}]{song2011}%
  \BibitemOpen
  \bibfield  {author} {\bibinfo {author} {\bibfnamefont {J.-W.}\ \bibnamefont
  {Song}}, \bibinfo {author} {\bibfnamefont {K.}~\bibnamefont {Yamashita}}, \
  and\ \bibinfo {author} {\bibfnamefont {K.}~\bibnamefont {Hirao}},\ }\href
  {\doibase 10.1063/1.3628522} {\bibfield  {journal} {\bibinfo  {journal} {J.\
  Chem.\ Phys.}\ }\textbf {\bibinfo {volume} {135}},\ \bibinfo {pages} {071103}
  (\bibinfo {year} {2011})}\BibitemShut {NoStop}%
\bibitem [{\citenamefont {Song}\ \emph {et~al.}(2013)\citenamefont {Song},
  \citenamefont {Giorgi}, \citenamefont {Yamashita},\ and\ \citenamefont
  {Hirao}}]{song2013}%
  \BibitemOpen
  \bibfield  {author} {\bibinfo {author} {\bibfnamefont {J.-W.}\ \bibnamefont
  {Song}}, \bibinfo {author} {\bibfnamefont {G.}~\bibnamefont {Giorgi}},
  \bibinfo {author} {\bibfnamefont {K.}~\bibnamefont {Yamashita}}, \ and\
  \bibinfo {author} {\bibfnamefont {K.}~\bibnamefont {Hirao}},\ }\href
  {\doibase 10.1063/1.4811775} {\bibfield  {journal} {\bibinfo  {journal} {J.\
  Chem.\ Phys.}\ }\textbf {\bibinfo {volume} {138}},\ \bibinfo {pages} {241101}
  (\bibinfo {year} {2013})}\BibitemShut {NoStop}%
\bibitem [{\citenamefont {Marzari}\ and\ \citenamefont
  {Vanderbilt}(1997)}]{PhysRevB.56.12847}%
  \BibitemOpen
  \bibfield  {author} {\bibinfo {author} {\bibfnamefont {N.}~\bibnamefont
  {Marzari}}\ and\ \bibinfo {author} {\bibfnamefont {D.}~\bibnamefont
  {Vanderbilt}},\ }\href {\doibase 10.1103/PhysRevB.56.12847} {\bibfield
  {journal} {\bibinfo  {journal} {Phys. Rev. B}\ }\textbf {\bibinfo {volume}
  {56}},\ \bibinfo {pages} {12847} (\bibinfo {year} {1997})}\BibitemShut
  {NoStop}%
\bibitem [{\citenamefont {Souza}\ \emph {et~al.}(2001)\citenamefont {Souza},
  \citenamefont {Marzari},\ and\ \citenamefont
  {Vanderbilt}}]{PhysRevB.65.035109}%
  \BibitemOpen
  \bibfield  {author} {\bibinfo {author} {\bibfnamefont {I.}~\bibnamefont
  {Souza}}, \bibinfo {author} {\bibfnamefont {N.}~\bibnamefont {Marzari}}, \
  and\ \bibinfo {author} {\bibfnamefont {D.}~\bibnamefont {Vanderbilt}},\
  }\href {\doibase 10.1103/PhysRevB.65.035109} {\bibfield  {journal} {\bibinfo
  {journal} {Phys. Rev. B}\ }\textbf {\bibinfo {volume} {65}},\ \bibinfo
  {pages} {035109} (\bibinfo {year} {2001})}\BibitemShut {NoStop}%
\bibitem [{\citenamefont {Mostofi}\ \emph {et~al.}(2008)\citenamefont
  {Mostofi}, \citenamefont {Yates}, \citenamefont {Lee}, \citenamefont {Souza},
  \citenamefont {Vanderbilt},\ and\ \citenamefont {Marzari}}]{mostofi2008}%
  \BibitemOpen
  \bibfield  {author} {\bibinfo {author} {\bibfnamefont {A.~A.}\ \bibnamefont
  {Mostofi}}, \bibinfo {author} {\bibfnamefont {J.~R.}\ \bibnamefont {Yates}},
  \bibinfo {author} {\bibfnamefont {Y.-S.}\ \bibnamefont {Lee}}, \bibinfo
  {author} {\bibfnamefont {I.}~\bibnamefont {Souza}}, \bibinfo {author}
  {\bibfnamefont {D.}~\bibnamefont {Vanderbilt}}, \ and\ \bibinfo {author}
  {\bibfnamefont {N.}~\bibnamefont {Marzari}},\ }\href {\doibase
  10.1016/j.cpc.2007.11.016} {\bibfield  {journal} {\bibinfo  {journal}
  {Comput.\ Phys.\ Commun.}\ }\textbf {\bibinfo {volume} {178}},\ \bibinfo
  {pages} {685} (\bibinfo {year} {2008})}\BibitemShut {NoStop}%
\bibitem [{\citenamefont {Mu}\ \emph {et~al.}(2022)\citenamefont {Mu},
  \citenamefont {Wang}, \citenamefont {Varley}, \citenamefont {Lyons},
  \citenamefont {Wickramaratne}, ,\ and\ \citenamefont {{Van de
  Walle}}}]{Mu_2022}%
  \BibitemOpen
  \bibfield  {author} {\bibinfo {author} {\bibfnamefont {S.}~\bibnamefont
  {Mu}}, \bibinfo {author} {\bibfnamefont {M.}~\bibnamefont {Wang}}, \bibinfo
  {author} {\bibfnamefont {J.~B.}\ \bibnamefont {Varley}}, \bibinfo {author}
  {\bibfnamefont {J.~L.}\ \bibnamefont {Lyons}}, \bibinfo {author}
  {\bibfnamefont {D.}~\bibnamefont {Wickramaratne}}, , \ and\ \bibinfo {author}
  {\bibfnamefont {C.~G.}\ \bibnamefont {{Van de Walle}}},\ }\href {\doibase
  10.1103/PhysRevB.105.155201} {\bibfield  {journal} {\bibinfo  {journal}
  {Phys.\ Rev.\ B}\ }\textbf {\bibinfo {volume} {105}},\ \bibinfo {pages}
  {155201} (\bibinfo {year} {2022})}\BibitemShut {NoStop}%
\bibitem [{Foo()}]{Footnote2}%
  \BibitemOpen
  \href@noop {} {}\bibinfo {note} {Thermo\_pw code is available from
  https://dalcorso.github.io/thermo\_pw/}\BibitemShut {NoStop}%
\bibitem [{\citenamefont {Sturm}\ \emph {et~al.}(2015)\citenamefont {Sturm},
  \citenamefont {Furthm{\"u}ller}, \citenamefont {Bechstedt}, \citenamefont
  {Schmidt-Grund},\ and\ \citenamefont {Grundmann}}]{Sturm_2015}%
  \BibitemOpen
  \bibfield  {author} {\bibinfo {author} {\bibfnamefont {C.}~\bibnamefont
  {Sturm}}, \bibinfo {author} {\bibfnamefont {J.}~\bibnamefont
  {Furthm{\"u}ller}}, \bibinfo {author} {\bibfnamefont {F.}~\bibnamefont
  {Bechstedt}}, \bibinfo {author} {\bibfnamefont {R.}~\bibnamefont
  {Schmidt-Grund}}, \ and\ \bibinfo {author} {\bibfnamefont {M.}~\bibnamefont
  {Grundmann}},\ }\href {\doibase 10.1063/1.4934705} {\bibfield  {journal}
  {\bibinfo  {journal} {APL Materials}\ }\textbf {\bibinfo {volume} {3}},\
  \bibinfo {pages} {106106} (\bibinfo {year} {2015})}\BibitemShut {NoStop}%
\bibitem [{\citenamefont {Furthm\"uller}\ and\ \citenamefont
  {Bechstedt}(2016)}]{Furthmuller_2016}%
  \BibitemOpen
  \bibfield  {author} {\bibinfo {author} {\bibfnamefont {J.}~\bibnamefont
  {Furthm\"uller}}\ and\ \bibinfo {author} {\bibfnamefont {F.}~\bibnamefont
  {Bechstedt}},\ }\href {\doibase 10.1103/PhysRevB.93.115204} {\bibfield
  {journal} {\bibinfo  {journal} {Phys. Rev. B}\ }\textbf {\bibinfo {volume}
  {93}},\ \bibinfo {pages} {115204} (\bibinfo {year} {2016})}\BibitemShut
  {NoStop}%
\bibitem [{\citenamefont {Fujiwara}(2007)}]{Fujiwara_2007}%
  \BibitemOpen
  \bibfield  {author} {\bibinfo {author} {\bibfnamefont {H.}~\bibnamefont
  {Fujiwara}},\ }\href@noop {} {\emph {\bibinfo {title} {Spectroscopic
  {Ellipsometry}}}}\ (\bibinfo  {publisher} {John Wiley \& Sons},\ \bibinfo
  {address} {New York},\ \bibinfo {year} {2007})\BibitemShut {NoStop}%
\end{thebibliography}
\end{document}